\documentclass[iop,apj]{emulateapj}
\usepackage{apjfonts}
\begin{document}
\newcommand{\comment}[1]{}

\newcommand{\Msun}{M_{\odot}}
\newcommand{\mvir}{M_\mathrm{vir}}
\newcommand{\vmax}{V_\mathrm{max}}
\newcommand{\vmac}{V_\mathrm{max}^\mathrm{acc}}
\newcommand{\mvac}{M_\mathrm{vir}^\mathrm{acc}}
\newcommand{\sfr}{\mathrm{SFR}}
\newcommand{\plotgrace}[1]{\includegraphics[angle=-90,width=\columnwidth,type=eps,ext=.eps,read=.eps]{#1}}
\newcommand{\plotgraceflip}[1]{\includegraphics[angle=-90,width=\columnwidth,type=eps,ext=.eps,read=.eps]{#1}}
\newcommand{\plotlargegrace}[1]{\includegraphics[angle=-90,width=2\columnwidth,type=eps,ext=.eps,read=.eps]{#1}}
\newcommand{\plotlargegraceflip}[1]{\includegraphics[angle=-90,width=2\columnwidth,type=eps,ext=.eps,read=.eps]{#1}}
\newcommand{\plotminigrace}[1]{\includegraphics[angle=-90,width=0.5\columnwidth,type=eps,ext=.eps,read=.eps]{#1}}
\newcommand{\plotmicrograce}[1]{\includegraphics[angle=-90,width=0.25\columnwidth,type=eps,ext=.eps,read=.eps]{#1}}
\newcommand{\plotsmallgrace}[1]{\includegraphics[angle=-90,width=0.66\columnwidth,type=eps,ext=.eps,read=.eps]{#1}}

\shortauthors{BEHROOZI, CONROY \& WECHSLER} 
\shorttitle{Uncertainties Affecting the Stellar Mass -- Halo Mass
  Relation}

\title{A Comprehensive Analysis of Uncertainties Affecting the\\Stellar
  Mass -- Halo Mass Relation for $0<\lowercase{z}<4$}

\author{Peter S. Behroozi\altaffilmark{1}, Charlie
  Conroy\altaffilmark{2}, Risa H. Wechsler\altaffilmark{1}}
\altaffiltext{1}{Kavli Institute for Particle Astrophysics and
  Cosmology; Physics Department, Stanford University, and SLAC National
  Accelerator Laboratory, Stanford, CA 94305}
\altaffiltext{2}{Department of Astrophysical Sciences, Princeton
  University, Princeton, NJ 08544}

\begin{abstract}
  We conduct a comprehensive analysis of the relationship between
  central galaxies and their host dark matter halos, as characterized
  by the stellar mass -- halo mass (SM--HM) relation, with rigorous
  consideration of uncertainties.  Our analysis focuses on results
  from the abundance matching technique, which assumes that every dark
  matter halo or subhalo above a specific mass threshold hosts one
  galaxy.  We provide a robust estimate of the SM--HM relation for
  $0<z<1$ and discuss the quantitative effects of uncertainties in
  observed galaxy stellar mass functions (GSMFs) (including stellar
  mass estimates and counting uncertainties), halo mass functions
  (including cosmology and uncertainties from substructure), and the
  abundance matching technique used to link galaxies to halos
  (including scatter in this connection).  Our analysis results in a
  robust estimate of the SM--HM relation and its evolution from z=0 to
  z=4.  The shape and evolution are well constrained for $z < 1$.  The
  largest uncertainties at these redshifts are due to stellar mass
  estimates (0.25 dex uncertainty in normalization); however, failure
  to account for scatter in stellar masses at fixed halo mass can lead
  to errors of similar magnitude in the SM--HM relation for central
  galaxies in massive halos.  We also investigate the SM--HM relation
  to $z=4$, although the shape of the relation at higher redshifts
  remains fairly unconstrained when uncertainties are taken into
  account.  We find that the integrated star formation at a given halo
  mass peaks at 10-20\% of available baryons for all redshifts from 0
  to 4.  This peak occurs at a halo mass of $7 \times 10^{11} \Msun$
  at $z=0$ and this mass increases by a factor of 5 to $z = 4$.  At
  lower and higher masses, star formation is substantially less
  efficient, with stellar mass scaling as $M_\ast \sim M_h^{2.3}$ at
  low masses and $M_\ast \sim M_h ^{0.29}$ at high masses.  The
  typical stellar mass for halos with mass less than $10^{12}$ $\Msun$
  has increased by $0.3-0.45$ dex for halos since $z \sim 1$.  These
  results will provide a powerful tool to inform galaxy evolution
  models.
\end{abstract}

\section{Introduction}

A variety of physical processes are thought to be responsible for the
observed distribution of galaxy properties, and distinguishing among
them is one of the principal goals of modern galaxy formation theory.
Among the relevant mechanisms are those responsible for galaxy
growth, such as star formation and galaxy mergers, as well as those
responsible for regulating growth, including energetic feedback by
supernovae, active galactic nuclei, cosmic ray pressure, 
and long gas cooling times.

A fruitful approach to separating the influence of different
mechanisms is to constrain the redshift--dependent relation between
physical characteristics of galaxies, such as stellar mass, and the
mass of their dark matter halos.  This is possible because it is
expected that many of these physical processes depend primarily on the
mass of a galaxy's dark matter halo.  By connecting galaxies to their
parent halos, one is able to more clearly identify and constrain the
physical processes responsible for galaxy growth.

The galaxy stellar mass -- halo mass relation has additional utility
because many properties of both galaxies and halos are tightly
correlated with halo mass.  In addition, the stellar mass -- halo mass
relation provides a mechanism for connecting predictions for the halo
mass function and the mass-dependent spatial clustering of halos to
the abundances and clustering of galaxies.  If a model for galaxy
evolution is able to reproduce the intrinsic galaxy mass -- halo mass
relation in the correct cosmological model, then such a model will
match both the observed stellar mass function and the stellar mass
dependent clustering of galaxies.  Simultaneously matching these two
observational quantities and their evolution has been difficult with
either hydrodynamical simulations or semi-analytic models of galaxy
formation \citep[e.g.,][]{Weinberg04,Li07}.

There are several ways to constrain the galaxy mass -- halo
mass relation.  The first type of approach attempts to directly
measure the mass of galactic halos.  Techniques include weak lensing
\citep[e.g.,][]{Guzik02, Sheldon04, mandelbaum-06} and the use of
satellite galaxy or stellar velocities as tracers of the halo potential well
\citep[e.g.,][]{Ashman93, Zaritsky94, Prada03, VDB04b, conroy-07}.  While such
methods are a relatively direct probe of halo mass, they are limited
in dynamic range; current observations probe halo masses from roughly
$10^{12}$--$10^{14}$ $\Msun$ at the present epoch, and a smaller range
at higher redshift.  A second approach is to identify groups and
clusters of galaxies either through optical or X-ray selected cluster
catalogs, and then directly measure their galaxy content
\citep[e.g.,][]{LinMohr04, Hansen09, Yang07}.  This is limited to
relatively massive halos (although for optically identified groups it
could extend to lower masses as new surveys probe dimmer galaxies in
large enough volumes), and it also requires accurate knowledge of the
mass--observable relation \citep{Yang07,Hansen09}.

An alternative approach is to assume that the properties of the halo
population are known, for example from cosmological simulations, and
then find a functional form relating galaxies to halos which achieves
agreement with a variety of observations.  This approach is
less direct but can be applied over a much larger dynamic range.  Halo
occupation \citep[e.g.,][]{Berlind02, Bullock02, Cooray02, Tinker05,
  zheng-07} and conditional luminosity function modeling
\citep[e.g.,][]{Yang03, Cooray06} fall into this category.

In the past decade, a number of studies have found that this latter
approach can be greatly simplified using a technique called abundance
matching.  In its most basic form, the technique assigns the most
massive (or the most luminous) galaxies to the most massive halos
monotonically.  The technique thus requires as input only the observed
abundance of galaxies as a function of mass, namely the galaxy stellar
mass function (alternatively the galaxy luminosity function) and the
abundance of dark matter halos as a function of mass, namely the halo
mass function.  This technique has been shown to accurately reproduce
a variety of observational results including various measures of the
redshift-- and scale--dependent spatial clustering of galaxies
\citep{Colin99, kravtsov_klypin:99, Neyrinck04, kravtsov:04, Vale04,
  Vale06, Tasitsiomi04, conroy:06, Shankar06, Berrier06, Marin08,
  Guo-09, moster-09}. In the context of this technique, not
only central halos but also subhalos (halos contained within the
virial radii of larger halos) are included in the matching process,
meaning that satellite galaxies can be accounted for without any
additional parameters.

Applications of the abundance matching technique have typically
focused on using the default modeling assumptions to derive
statistical information about the galaxy -- halo connection.  Uncertainties in
the derived galaxy mass -- halo mass relation have received little
systematic attention in the context of this technique \citep[though
see][for a recent treatment of the attendant
uncertainties]{moster-09}.  An accurate assessment of the
uncertainties is necessary to make strong statements regarding the
underlying physical processes responsible for the derived galaxy -- halo
relation.  In the present work we undertake an exhaustive exploration
of the uncertainties relevant in constructing the galaxy stellar mass
-- halo mass relation from the abundance matching technique.  We
consider a range of uncertainties related to the observational stellar
mass function, the theoretical halo mass function, and the underlying
technique of abundance matching.  The resulting galaxy stellar mass --
halo mass relation and associated uncertainties will provide a
benchmark against which galaxy evolution models may be fruitfully
tested.

This paper is divided into several sections.  In \S\ref{s:uncertainties} we detail
known sources of uncertainty which may affect our results.  Our
methodology for modeling the effects of uncertainties is discussed in
\S \ref{s:methodology}, providing simple conversions where possible to
allow for different modeling choices.  We present our resulting
estimates of the galaxy stellar mass -- halo mass relation for $z<1$
in \S \ref{s:results} and describe the contribution of each of the
uncertainties to the overall error budget.  Estimates for the
evolution of the relation out to $z\sim4$, for which the uncertainties
are significantly less well-understood, are presented in \S
\ref{s:results_highz}.  Finally, we discuss the implications of this
work in \S \ref{s:discussion} and summarize our conclusions in \S
\ref{s:conclusions}.

Stellar masses throughout are quoted assuming a
\citet{chabrier-2003-115} initial mass function (IMF), the stellar population
synthesis models of \citet{bc-03}, and the age and dust models in
\citet{blanton-roweis-07}.
We consider multiple cosmologies in this paper, but the main results assume a
WMAP5+SN+BAO concordance $\Lambda$CDM cosmology \citep{wmap5} with
$\Omega_M = 0.27$, $\Omega_\Lambda = 1 - \Omega_M$, $h = 0.7$,
$\sigma_8 = 0.8$, and $n_s = 0.96$.

\section{Uncertainties Affecting the Stellar Mass to Halo Mass
  Relation}
\label{s:uncertainties}

Uncertainties in the abundance matching technique for assigning
galaxies to dark matter halos can be conceptually separated into three
classes.  The first is uncertainty in the abundance of galaxies as a
function of stellar mass.  This class includes both uncertainties in
counting galaxies due to shot noise and sample variance, as well as
uncertainties in the stellar mass estimates themselves.  The second class
concerns the dark matter halos, and includes uncertainties in
cosmological parameters, the impact of baryon condensation, and
substructure.  Finally, there are uncertainties in the process of
matching galaxies to halos arising primarily from the intrinsic
scatter between galaxy stellar mass and halo mass.  Each of these
sources of uncertainty are described in detail below.  The detailed
modeling of these uncertainties is described in \S
\ref{s:methodology}.

\subsection{Uncertainties in the Stellar Mass Function}

Galaxy stellar masses are not measured directly, but are instead
inferred from photometry and/or spectra.  In particular, as the
observed stellar light is a function of many physical processes (e.g.,
stellar evolution, star formation history and metal--enrichment
history, and wavelength--dependent dust attenuation), stellar masses
are estimated via complicated models to find the best fit to galaxy
observations in a very large parameter space.  Assumptions and
simplifications in these models, along with the fact that the best fit
may be only one of a number of likely possibilities, mean that there
can be substantial uncertainties in calculated stellar masses.

Different types of observations can yield different uncertainties in
these calculations.  Spectroscopic surveys generally recover more
spectral features per galaxy and more accurate redshifts than
photometric surveys.  However, spectroscopy requires substantially
more telescope time than photometry.  Therefore, spectroscopic samples
tend to be limited both in area and depth, which translates into
limitations in both volume and the minimum stellar mass probed.  An
additional problem for spectroscopic surveys such as the Sloan Digital
Sky Survey \citep[SDSS,][]{York00} is that the spectra probe only the
central regions of galaxies (the SDSS spectra gather on average $1/3$
of the total flux from galaxies at $z=0.1$).  For galaxies containing
both a bulge and a disk, or galaxies with radial gradients, the
spectra will therefore not provide a fair sample of the entire galaxy
\citep[e.g.,][]{Kewley05}; furthermore, this bias will be a function
of redshift.  For these reasons, especially for robust comparisons of
evolution, we confine this paper to photometric--based stellar masses;
however, samples with spectroscopic redshifts are used where
available.

\subsubsection{Principal Uncertainties in Stellar Mass Functions}

\label{syst_stellarm}
In this paper, we analyze seven main sources of uncertainties in
stellar mass functions applicable to photometric surveys, listed below
in rough order of importance.

\begin{enumerate}
\item {\em Choice of stellar initial mass function (IMF):} The
  luminosity of stars scales as their mass to a large power (${\rm
    dln}L/{\rm dln}M\sim3.5$), while the IMF, defined as the number of
  stars formed per unit mass, scales approximately as $\xi\propto
  M^{-2.3}$ \citep{Salpeter55}.  Thus, the light from a galaxy is
  dominated by the most massive stars, while the total stellar mass is
  dominated by the lowest mass stars.  This fact, which has been known
  for decades \citep[e.g.,][]{Tinsley76}, implies that the assumed
  form of the IMF at $M\lesssim0.5\Msun$ will have no effect on the
  integrated light from galaxies, but will have a large effect on the
  total stellar mass.  Nonetheless, constraints on the total dynamical
  mass of spheroidal systems provide a valuable independent check on
  the form of the IMF at low masses.  \citet{Cappellari06} find, for
  example, that the IMF proposed by \citet{chabrier-2003-115} for the
  Solar neighborhood is consistent with dynamical constraints on
  masses of nearby ellipticals.

  We do not marginalize over IMFs in our error calculations for the
  reason that it is relatively simple to convert from our choice of
  IMF \citep{chabrier-2003-115} to other IMFs.  For our purposes the
  IMF becomes a serious source of systematic uncertainty only if the
  IMF varies with galaxy properties or if it evolves. It should be
  noted, however, that the IMF can also introduce more complicated
  systematic effects associated with the inferred star formation rate,
  which may in turn impact stellar masses in non--trivial ways
  \citep[e.g.][]{conroy-09}.

\item {\em Choice of Stellar Population Synthesis (SPS) model:} SPS
  modeling efforts have grown substantially in sophistication in the
  past decade \citep[e.g.][]{Leitherer99, bc-03, LeBorgne04,
    maraston-2005}.  Yet, significant uncertainties remain
  \citep[e.g.,][]{Charlot96a, Charlot96b, Yi03, LeeHC07, Conroy09}.
  Treatments of convection vary, leading to different main sequence
  turn off times for intermediate mass stars.  Advanced stages of
  stellar evolution, including blue stragglers, thermally--pulsating
  AGB stars (TP--AGB), and horizontal branch stars, are poorly
  understood, both observationally and theoretically.  The theoretical
  spectral libraries contain known deficiencies, especially for M
  giants, where the effective temperatures are low and where
  hydrodynamic effects become important.  Empirical stellar libraries
  to some extent circumvent these issues, although they are plagued by
  incomplete coverage in the HR diagram and difficulties associated
  with deriving stellar parameters.  See \citet{Conroy09} and
  \citet{Percival09b} for recent reviews.

  Different stellar population synthesis models treat these issues
  differently, which can result in large systematic differences in the
  derived stellar mass.  For instance, the model of
  \cite{maraston-2005} compared to \cite{bc-03} has systematic
  differences of 0.1dex in stellar mass
  \citep{salimbeni-09,perezgonzalez-2007}.  However,
  \cite{salimbeni-09} reports that the model in \cite{cb-07} (with a
  revised treatment of TP-AGB stars) yields systematic differences in
  stellar mass relative to \cite{bc-03}, which ranges from 0.05dex for
  $10^{11}$$\Msun$ galaxies to 0.3dex for $10^{9.5}$$\Msun$ galaxies.
  \citet{Conroy09} show that uncertainty in the luminosity of the
  TP-AGB phase can shift stellar masses by as much as $\pm0.2$ dex.

\item {\em Parameterization of star formation histories:} In order to
  estimate stellar masses, model libraries are constructed with a
  large range in star formation histories (SFHs), dust attenuation
  (see below), and, often but not always, metallicity.  The adopted
  functional form of the SFH is another source of systematic
  uncertainty, as typically very simple functional forms are assumed.
  Several authors have investigated various aspects of this problem.
  When attempting to model observed photometry,
  \cite{perezgonzalez-2007} found that a single stellar population
  model (in particular, star formation proportional to $e^{-t/\tau}$,
  which is a commonly-used parameterization) systematically
  underpredicts stellar mass by 0.18 dex compared to a double stellar
  population model (exponentially decaying star formation followed by
  a later starburst).  The particular parameterization of SFHs may
  also lead to systematic differences as a function of stellar mass.
  \citet{LeeSK08} analyzed a sample of mock Lyman--break galaxies at
  $z\sim4-5$ and found that simple SFHs produced best--fit stellar
  masses that were under or overestimated by $\sim\pm50$\% depending
  on the rest-frame galaxy color.  This bias was attributed to the
  chaotic SFHs of the mock galaxies.

\item {\em Choice of dust attenuation model:} Because dust reddens
  starlight, it is difficult to separate the effects of dust from
  stellar population effects, especially when fitting optical
  photometry.  The effects of dust are also known  to change depending
  on galaxy inclination \citep[e.g.,][]{Driver07}.  Hence, the choice of dust
  attenuation law has a
  nontrivial effect on the inferred stellar population ages and,
  consequently, star formation histories derived from photometric and
  even spectroscopic surveys \citep{panter:2007}.  In terms of the
  effect on stellar masses, \cite{perezgonzalez-2007} compared the
  dust models of \cite{calzetti-00} and \cite{cf-00}, finding a
  systematic difference of 0.10dex. \cite{panter:2007} found a similar
  difference between Calzetti et al.\ and models based on extinction
  curves from the Small and Large Magellanic Clouds.  The effects of
  varying dust attenuation models have also been explored recently by
  \citet{marchesini-2008} and \citet{Muzzin09}, with similar results.
  We have used the stellar population fitting procedure described in
  \citet{Conroy09} to compare the Calzetti et al.\ dust attenuation law
  to the dust model used in the \texttt{kcorrect} package
  \citep{blanton-roweis-07}.  We find a median offset of 0.02dex but
  also a systematic trend such that two galaxies whose stellar masses
  are estimated with Calzetti dust attenuation and are separated by
  1.0dex will have \texttt{kcorrect} masses separated by (on average)
  only 0.92dex.

\item {\em Statistical errors in individual stellar mass estimates:}
  Stellar mass estimates for each galaxy are subject to statistical
  errors due to uncertainty in photometry as well as uncertainty in
  the SPS parameters for a given set of model assumptions.  We treat
  this here as a random statistical error.  While it may seem that
  random scatter in individual stellar masses should on average have
  no systematic effect, it in fact introduces a systematic error
  analogous to Eddington bias \citep{Eddington40} observed in
  luminosity functions.  As the stellar mass function drops off
  steeply beyond a certain characteristic stellar mass, there are many
  more low stellar mass galaxies that can be up-scattered than there
  are high stellar mass galaxies that can be down-scattered by errors
  in stellar mass estimates.  This asymmetric scatter implies that the
  drop-off in number density at high masses becomes shallower in the
  presence of scatter.  We discuss this effect in detail in
  $\S$\ref{measurement_scatter} \citep[see also the Appendix
  in][]{Cattaneo08}.

\item {\em Sample variance:} Surveys of finite regions of the Universe
  are susceptible to large--scale fluctuations in the number density
  of galaxies.  This is no longer a dominant source of uncertainty for
  the volumes probed at low redshift by the SDSS, but it is an
  important consideration for higher--redshift surveys, which cover
  much smaller comoving volumes.  Most authors who consider sample
  variance attempt to minimize it by averaging over several fields
  \citep[e.g.,][]{perezgonzalez-2007, marchesini-2008}.  Very few
  surveys at $z>0$ attempt to estimate the magnitude of the error
  except by computing the field--to--field variance, which is often an
  underestimate when insufficient volume is probed \citep{Crocce09}.  We
  detail a more accurate method based on simulations to model the
  error arising from sample variance in $\S$\ref{sample_variance}.

\item {\em Redshift errors:} Photometric redshift errors blur the
  distinction between GSMFs at different redshifts.  While a galaxy
  may be scattered either up or down in redshift space, volume-limited
  survey lightcones will contain larger numbers of galaxies at higher
  redshifts, meaning that the GSMF as reported at lower redshifts will
  be artificially inflated.  Moreover, as galaxies at earlier times
  have lower stellar masses, surveys will tend to report artificially
  larger faint-end slopes in the GSMF.  However, as these errors are
  well known, it is easy to correct for their effects on the stellar
  mass function, as has been done for the data in
  \cite{perezgonzalez-2007} (see the appendix of
  \citealt{perezgonzalez-05} for details on this process).

\end{enumerate}

For completeness, we remark that galaxy-galaxy lensing will also
result in systematic errors in the GSMF at high redshifts because
galaxy magnification will result in higher observed luminosities.
However, from ray-tracing studies of the Millennium simulation
\citep{hilbert-07}, the expected scatter in galaxy stellar masses from
lensing is minimal (e.g., 0.04 dex at $z=1$) compared to the other
sources of scatter above (e.g., 0.25 dex from different model
choices).  For that reason, we do not model galaxy-galaxy lensing
effects in this paper.

\subsubsection{Additional Systematics at $z>1$}
\label{SFRs}

Recently, it has become clear that current estimates of the evolution
in the cosmic SFR density are not consistent with estimates of the
evolution of the stellar mass density at $z>1$ \citep{Nagamine06,
  Hopkins06b, perezgonzalez-2007, Wilkins08}.  The origin of this
discrepancy is currently a matter of debate.  One solution involves
allowing for an evolving IMF with redshift \citep{Dave08, Wilkins08}.
While such a solution is controversial, a number of independent lines
of evidence suggest that the IMF was different at high redshift
\citep{Lucatello05, Tumlinson07a, Tumlinson07b, vanDokkum08}.
\citet{Reddy09} offer a more mundane explanation for the discrepancy.
They appeal to luminosity--dependent reddening corrections in the
ultraviolet luminosity functions at high redshift, and demonstrate
that the purported discrepancy then largely vanishes.

In contrast to results at $z>1$, there does seem to be an accord that
for $z<1$ both the integrated SFR and the total stellar mass are in
good agreement if one assumes (as we have) a \citet{chabrier-2003-115}
IMF \citep[see][]{wilkins-2008,perezgonzalez-2007,hopkins-2006-651,
  nagamine-2006-653, cw-08}.

Because of the discrepancy between reported SFRs and stellar masses in
the literature, it is clear that estimates of uncertainties in galaxy
stellar mass functions and SFRs at $z>1$ tend to underestimate the
true uncertainties; for this reason, we separately analyze results for
$z<1$ in \S \ref{s:results} and $z>1$ in \S \ref{s:results_highz} of
this paper. 

\subsection{Uncertainties in the Halo Mass Function}

Dark matter halo properties over the mass range $10^{10}-10^{15}$
$\Msun$ have been extensively analyzed in simulations
\citep[e.g.,][]{Jenkins01, Warren06, tinker-umf}, and the overall
cosmology has been constrained by probes such as WMAP
\citep[][]{wmap1,wmap5}.  As such, uncertainties in the halo mass
function have on the whole much less impact than uncertainties in the
stellar mass function.  We present our primary results for a fixed
cosmology (WMAP5), but we also calculate the impact of uncertain
cosmological parameters on our error bars.  We do not marginalize over
the mass function uncertainties for a given cosmology, as the relevant
uncertainties are constrained at the 5\% level \citep[when baryonic
effects are neglected, see below;][]{tinker-umf}.  Additionally, in
Appendix \ref{direct_matching}, a simple method is described to
convert our results to a different cosmology using an arbitrary mass
function.  For completeness, we mention the three most significant
uncertainties here:

\begin{enumerate}
\item {\em Cosmological model:} The stellar mass -- halo mass relation
  has dependence on cosmological parameters due to the resulting
  differences in halo number densities.  We investigate this both by
  calculating the relation for two specific cosmological modes (WMAP1
  and WMAP5 parameters) and then by calculating the uncertainties in
  the relation over the full range of cosmologies allowed by WMAP5
  data.  We find that in all cases these uncertainties are small
  compared to the uncertainties inherent in stellar mass modeling (\S
  \ref{syst_stellarm}), although they are larger than the statistical
  errors for typical halo masses at low redshift.

\item {\em Uncertainties in substructure identification:} Different
  simulations have different methods of identifying and assigning
  masses to substructure.  Our matching methods make use only of the
  subhalo mass at the epoch of accretion ($M_\mathrm{acc}$) as this
  results in a better match to clustering and pair--count results
  \citep{conroy:06, Berrier06}, so we are largely immune to the
  problem of different methods for calculating subhalo masses.  Of
  greater concern is the ability to reliably follow subhalos in
  simulations as they are tidally stripped.  Two related issues apply
  here.  The first is that it is not clear how to account for subhalos
  which fall below the resolution limit of the simulation.  The second
  is that the formation of galaxies will dramatically increase the
  binding energy of the central regions of subhalos, potentially
  making them more resilient to tidal disruption.  Hydrodynamic
  simulations suggest that this latter effect is small except for
  subhalos that orbit near the centers of the most massive clusters
  \citep{Weinberg08}.  However, while these details are important for accurately
  predicting the clustering strength on small scales ($\lesssim1$
  Mpc), they are not a substantial source of uncertainty for the global
  halo mass --- stellar mass relation because satellites are always
  sub-dominant ($\lesssim$ 20\%) by number.  We discuss the analytic
  method we use to model the satellite contribution to the halo mass
  function in \S \ref{s:mass_functions}.

\item {\em Baryonic physics:} Recent work by \cite{Stanek09} suggests
  that gas physics can affect halo masses relative to dark matter-only
  simulations by -16\% to +17\%, leading to number density shifts of
  up to 30\% in the halo mass function at $10^{14}$ $\Msun$.  Without
  evidence for a clear bias in one direction or the other---the models
  of gas physics still remain too uncertain---we do not apply a
  correction for this effect in our mass functions.  Uncertainties of
  this magnitude are larger than the statistical errors in individual
  stellar masses at low redshift, but are still small in comparison
  to systematic errors in calculating stellar masses.
\end{enumerate}

For completeness, we note that the effects of sample variance on halo
mass functions estimated from simulations are small.  Current
simulations readily probe volumes of 1000 ($h^{-1}$ Mpc)$^3$
\citep{tinker-umf}, and so the effects of sample variance on the halo
mass function are dwarfed by the effects of sample variance on the
stellar mass function; we therefore do not analyze them separately in
this paper.

We also remark on the issue of mass definitions.  Although abundance
matching implies matching the most massive galaxies to the most
massive halos, there is little consensus on \textit{which} halo mass
definition to use, with popular choices being $M_{vir}$ (mass within
the virial radius), $M_{200}$
(mass within a sphere with mean density 200$\rho_\mathrm{crit}$), and
$M_{fof}$ (mass determined by a friends-of-friends particle linking
algorithm).  We choose $M_{vir}$ for this paper and note that the
largest effect of choosing another mass algorithm will be a purely
definitional shift in halo masses.  We expect that scatter between any
two of these mass definitions is degenerate with and smaller than
the amount of scatter in stellar
masses at fixed halo mass (the latter effect is discussed in \S \ref{abund_match_uncertainties}).

\subsection{Uncertainties in Abundance Matching}
\label{abund_match_uncertainties}

Finally, there are two primary uncertainties concerning the abundance
matching technique itself:

\begin{enumerate}
\item {\em Nonzero scatter in assigning galaxies to halos:} While host
  halo mass is strongly correlated with stellar mass, the correlation
  is not perfect.  At a given halo mass, the halo merger history,
  angular momentum properties, and cooling and feedback processes can
  induce scatter between halo mass and galaxy stellar mass.  This is
  expected to result in scatter in stellar of $\sim$0.1--0.2 dex at a
  given halo mass, see \S 3.3.1 for discussion.  The scatter between
  halo mass and stellar mass will have systematic effects on the mean
  relation for reasons analogous to those mentioned for statistical
  error in stellar mass measurements.  At the high mass end where both
  the halo and stellar mass functions are exponential, scatter in
  stellar mass at fixed halo mass (or vice versa) will alter the
  average relation because there are more low mass galaxies that are
  upscattered than high mass galaxies that are downscattered.

\item {\em Uncertainty in Assigning Galaxies to Satellite Halos:} It
  is not clear that the halo mass --- stellar mass relation should be
  the same for satellite and central galaxies.  Once a halo is
  accreted onto a larger halo, it starts to lose halo mass because of
  dynamical effects such as tidal stripping.  While stripping of the
  halo appears to be a relatively dramatic process
  \citep[e.g.,][]{Kravtsov04b}, the stripping of the stellar component
  probably does not occur unless the satellite passes very near to the
  central object because the stellar component is much more tightly
  bound than the halo.  It is clear from the observed color--density
  relation \citep{Dressler80, Postman84, Hansen09} that star formation
  in satellite galaxies must eventually cease with respect to galaxies
  in the field.  It is less clear how quickly star formation ceases,
  and whether or not there is a burst of star formation upon
  accretion.  All of these issues can potentially alter the relation
  between halo and stellar mass for satellites (although the modeling
  results of \citealt{Wang06} suggest that the halo--satellite
  relation is indistinguishable from the overall galaxy--halo
  relation).

\end{enumerate}

\section{Methodology}
\label{s:methodology}

Our primary goal is to provide a robust estimate of the stellar mass
-- halo mass relation over a significant fraction of cosmic time via
the abundance matching technique.  We aim to construct this relation
by taking into account all of the relevant sources of uncertainty.
This section describes in detail a number of aspects of our
methodology, including our approach for incorporating uncertainties in
the stellar mass function ($\S$\ref{s:sys}), a summary of the adopted
halo mass functions and associated uncertainties ($\S$\ref{s:sim}),
the uncertainties associated with abundance matching (\S
\ref{s:amatch}), our choice of functional form for the stellar mass --
halo mass relation, including a discussion of why certain functions
should be preferred over others ($\S$\ref{s:func}), and the Markov
Chain Monte Carlo parameter estimation technique ($\S$\ref{s:mcmc}).
For readers interested in the general outline of our process but not
the details, we conclude with a brief summary of our methodology (\S
\ref{s:summary}).

\subsection{Modeling Stellar Mass Function Uncertainties}
\label{s:sys}

As discussed in \S\ref{s:uncertainties}, there are several classes of
uncertainties affecting the way the stellar mass function is used in
the abundance matching process.  In this section, we discuss
systematic shifts in stellar mass estimates and the effects of
statistical errors on the stellar mass function.

\subsubsection{Modeling Systematic Shifts in Stellar Mass Estimates}
\label{systematics_sm}

Most studies on the GSMF report Schechter function fits as well as
individual data points; many also provide statistical errors.
However, even when systematic errors are reported (either in Schechter
parameters or at individual data points), the systematic error
estimates are of limited value unless one is also able to model shifts
in the GSMF caused by such errors.

Fortunately, based on the discussion in $\S$\ref{syst_stellarm}, there
seem to be two main classes of systematic errors causing shifts in the
GSMF:

\begin{enumerate}
\item Over/underestimation of all stellar masses by a constant factor
  $\mu$.  This appears to cover the majority of errors, including most
  differences in SPS modeling, dust attenuation assumptions, and
  stellar population age models.
\item Over/underestimation of stellar masses by a factor which depends
  linearly on the logarithm of the stellar mass (i.e., depends on a
  power of the stellar mass).  This covers the majority of the
  remaining discrepancies between different SPS models and different
  stellar age models.
\end{enumerate}

Both forms of error are modeled with the equation
\begin{equation}
\log_{10}\left(\frac{M_{\ast,\mathrm{meas}}}{M_{\ast,\mathrm{true}}}\right) = 
\mu + \kappa\,\log_{10}\left(\frac{M_{\ast,\mathrm{true}}}{M_0}\right).
\label{eq:systematics}
\end{equation}
Without loss of generality, we may take $M_0 = 10^{11.3}$$\Msun$
(the fixed point of the variation between the \citealt{cb-07} and
\citealt{bc-03} models found by \citealt{salimbeni-09}), allowing the
prior on $M_0$ to be absorbed into the prior on $\mu$.

For the prior on $\mu$, we consider four contributing sources of
uncertainty.  We adopt estimates of the uncertainty from the SPS model
($\approx$0.1dex), the dust model ($\approx$0.1dex), and assumptions
about the star formation history ($\approx$0.2dex) from
\cite{perezgonzalez-2007} as detailed in \S \ref{syst_stellarm}.
Additionally, we have the variation in $\kappa \, \log_{10}(M_0)$ (at
most 0.1dex, as $|\kappa|\lesssim 0.15$ --- see below).  Assuming that
these are statistically independent, they combine to give a total
uncertainty of 0.25dex, which is consistent with the accepted range
for systematic uncertainties in stellar mass
\citep{perezgonzalez-2007,kannappan-07, vanderwel-06,marchesini-2008}.
For lack of adequate information (i.e., different models) to infer a
more complicated distribution, we assume that $\mu$ has a Gaussian
prior.  As more studies of the overall systematic shift $\mu$ become
available, our assumptions for the prior on $\mu$ and the probability
distribution will likely need corrections.  We remark, however, that
our results can easily be converted to a different assumption for
$\mu$, as $\mu$ simply imparts a uniform shift in the intrinsic
stellar masses relative to the observed stellar masses.

For the prior on $\kappa$, the result of \cite{salimbeni-09} would
suggest $|\kappa| \lesssim 0.15$.  As mentioned in
$\S$\ref{syst_stellarm}, we found that $|\kappa| \approx 0.08$ between
the \cite{blanton-roweis-07} and \cite{calzetti-00} models for dust
attenuation.  \cite{li-2009} finds $|\kappa| \lesssim 0.10$ between
\cite{blanton-roweis-07} and \cite{bell:2003} stellar masses.  Without
a large number of other comparisons, it is difficult to robustly
determine the prior distribution for $\kappa$; however, motivated by
the results just mentioned, we assume that the prior on $\kappa$ is a
Gaussian of width 0.10 centered at 0.0.

We remark that some authors have considered much more complicated
parameterizations of the systematic error.  For example, \cite{li-2009}
considers a four-parameter hyperbolic tangent fit to differences in
the GSMF caused by different SPS models, as well as a five-parameter
quartic fit.  However, we do not consider higher-order models for
systematic errors for several reasons.  First, given that second- and
higher-order corrections will result only in very small corrections to
the stellar masses in comparison to the zeroth-order correction ($\mu
\approx 0.25$dex), the corrections will not substantially effect the
systematic error bars.  Second, we do not know of any studies which
would allow us to construct priors on the higher-order corrections.
Finally, with higher-order models, there is the serious danger of
over-fitting---that is, with very loose priors on systematic errors,
the best-fit parameters for the systematic errors will be influenced
by bumps and wiggles in the stellar mass function due to statistical
and sample variance errors.  Hence, the interpretive value of the
systematic errors becomes increasingly dubious with 
each additional parameter.

\subsubsection{Modeling Statistical Errors in Individual Stellar Mass
  Measurements}
\label{measurement_scatter}

\newcommand{\ptrue}{\phi_\mathrm{true}}
\newcommand{\pobs}{\phi_\mathrm{meas}}
\newcommand{\pspit}{\phi_\mathrm{spitzer}}

In addition to the systematic effects discussed in the previous
section, measurement of stellar masses is subject to statistical
errors.  Even for a fixed set of assumptions about the dust model, SPS
model, and the parameterization of star formation histories, stellar
masses will carry uncertainties because the mapping between
observables and stellar masses is not one-to-one.  This additional
source of uncertainty has unique effects on the GSMF.  Observers will
see an GSMF ($\pobs$) which is the true or ``intrinsic'' GSMF
($\ptrue$) convolved with the probability distribution function of the
measurement scatter.  For instance, if the scatter is uniform across
stellar masses and has the shape of a certain probability distribution
$P$, we have:
\begin{equation}
\label{scatter_conv_eqn}
\pobs(M) = \int_{-\infty}^\infty \ptrue(10^y) \, P\left(y - \log_{10}(M)\right) dy, 
\label{true_to_obs}
\end{equation}
where $y$ is the integration variable, in units of $\log_{10}$ mass.
As derived in Appendix \ref{scatter_effects}, the approximate effect
of the convolution is
\begin{equation}
\label{scatter_conv_effect}
\log_{10} \left(\frac{\pobs(M)}{\ptrue(M)}\right) \approx \frac{\sigma^2}{2}\ln(10)\,\left(\frac{d\log\ptrue(M)}{d\log M}\right)^2, 
\end{equation}
where $\sigma$ is the standard deviation of $P$.  That is to say, the
effect of the convolution depends strongly on the logarithmic slope of
$\ptrue$.  Where the slope is small (i.e., for low-mass galaxies),
there is almost no effect.  Above $10^{11}$ $\Msun$,
where the GSMF becomes exponential, there can be a dramatic effect,
with the result that $\ptrue$ is more than an order of magnitude less than
$\pobs$ because it becomes far more likely that stellar mass
calculation errors produce a galaxy of very high perceived stellar
mass than it is for there to be such a galaxy in reality
\citep[see for example][]{Cattaneo08}.

For the observed $z\sim0$ GSMF, we take the probability distribution
$P$ to be log-normal with $1\sigma$ width $0.07$dex from the analysis
of the photometry of low--redshift luminous red galaxies (LRGs)
\citep{Conroy09}.  \citet{Kauffmann03a} found similar results
regarding the width of $P$.  This function only accounts for the
statistical uncertainties mentioned above and does not include additional
systematic uncertainties.  In light of Equation
\ref{scatter_conv_effect}, we use LRGs to estimate $P$ because LRGs
occupy the high stellar mass regime where measurement errors are most
likely to affect the shape of the observed GSMF.  However, the single
most important attribute of the distribution $P$ is its width; the
main results do not change substantially if an alternate distribution
with non-Gaussian tails beyond the $1\sigma$ limits of $P$ is used.

For higher redshifts, we scale the width of the probability
distribution to account for the fact that mass estimates become less
certain at higher redshift \citep[e.g.,][]{Conroy09, Kajisawa09}:

\begin{equation}
P(\Delta \log_{10} M_\ast, z) = \frac{\sigma_0}{\sigma(z)} 
	P_0\left(\frac{\sigma_0}{\sigma(z)} \Delta \log_{10} M_\ast\right),
\end{equation}
where $P_0$ is the probability distribution at $z=0$ (as discussed
above), $\sigma_0$ is the standard deviation of $P_0$, and $\sigma(z)$
gives the evolution of the standard deviation as a function of
redshift.

\cite{Conroy09} did not give a functional form for $\sigma(z)$, 
but they calculate for a handful of massive galaxies that
$\sigma(z=2)$ is $\approx 0.18$dex, as compared to $\sigma(z=0)
\approx$ 0.07dex. \cite{Kajisawa09} performed a similar calculation
(albeit with a different SPS model) of the distribution in several
redshift bins; their results show gradual evolution for $\sigma(z)$
out to $z=3.5$ for high stellar mass galaxies
consistent with a linear fit:
\begin{equation}
\label{sigma_evolution}
\sigma(z) = \sigma_0 + \sigma_z z.
\end{equation}
The results of \cite{Kajisawa09} suggest that $\sigma_z =$0.03-0.06
dex for LRGs.  As this is consistent with the value of $\sigma_z =
0.05$ dex which would correspond to \cite{Conroy09}, we adopt the
linear scaling of Equation \ref{sigma_evolution} with a Gaussian prior
of $\sigma_z=0.05\pm 0.015$ dex.

Note that the effect of this statistical error on the stellar mass
function is minimal below $10^{11}$ $\Msun$, and therefore does not
affect the stellar mass -- halo mass relation for halos below
$\sim10^{13}$ $\Msun$, as discussed in $\S$\ref{smmr_results}.
While this scatter does have an effect on the shape of the stellar
mass function for high-mass galaxies, the qualitative predictions we
make from this analysis are generic to all types of random scatter.

\subsection{Halo Mass Functions}
\label{s:sim}

The halo mass function specifies the abundance of halos as a function
of mass and redshift.  A number of analytic models and
simulation--based fitting functions have been presented for computing
mass functions given an input cosmology
\citep[e.g.,][]{PressSchechter74, Jenkins01, Warren06, tinker-umf}.
For most of our results we will adopt the universal mass function of
\citet{tinker-umf}, as described below.  Analytic mass functions are
preferable as they 1) allow mass functions to be computed for a range
of cosmologies and 2) do not suffer significantly from sample variance
uncertainties, because the analytic relations are typically calibrated
with very large or multiple $N-$body simulations.

For some purposes it will be useful to also consider full halo merger
trees derived directly from $N-$body simulations that have sufficient
resolution to follow halo substructures.  The simulations used herein
will be described below, in addition to our methods for modeling
uncertainties in the underlying mass function, including cosmology
uncertainties, sample variance in the galaxy surveys, and our models
for satellite treatment.

\subsubsection{Simulations}
\label{simulation}

For the principal simulation in this study (``L80G''), we used a pure
dark matter N-body simulation based on Adaptive Refinement Tree (ART)
code \citep{kravtsov_etal:97,kravtsov_klypin:99}.  The simulation
assumed flat, concordance $\Lambda$CDM ($\Omega_M = 0.3$,
$\Omega_\Lambda = 0.7$, $h = 0.7$, and $\sigma_8 = 0.9$) and included
$512^3$ particles in a cubic box with periodic boundary conditions and
comoving side length $80 h^{-1}$Mpc.  These parameters correspond to a
particle mass resolution of $\approx 3.2\times 10^8 h^{-1}$$\Msun$.
For this simulation, the ART code begins with a spatial grid size of
$512^3$; it refines the grid up to eight times in locally dense
regions, leading to an adaptive distance resolution of $\approx 1.2
h^{-1}$kpc (comoving units) in the densest parts and $\approx
0.31h^{-1}$Mpc in the sparsest parts of the simulation.

In this simulation, halos and subhalos were identified using a variant
of the Bound Density Maxima algorithm \citep{Klypin99}.  
Halo centers are located at peaks in the density field smoothed over a
24-particle SPH kernel (for a minimum resolvable halo mass of
$7.7\times 10^9 h^{-1}$ $\Msun$).  Nearby particles are classified
as bound or unbound in an iterative process; once all the locally
bound particles have been found, halo parameters such as the virial
mass $\mvir$ and maximum circular velocity $\vmax$ may be calculated.
(See \citealt{kravtsov:04} for complete details on the algorithm).
The simulation is complete down to $\vmax\approx 100$ km s$^{-1}$,
corresponding to a galaxy stellar mass of $10^{8.75}$ $\Msun$ at
$z=0$.

The ability of L80G to track satellites with high mass and force
resolution gives it several uses.  Merger trees from L80G inform our
prescription for converting analytical central-only halo mass
functions to mass functions which include satellite halos (see \S
\ref{s:mass_functions}).  Additionally, the merger trees allow for
evaluation of different models of satellite stellar evolution with
full consistency (see \S \ref{s:sats}).  Finally, the knowledge of
which satellite halos are associated with which central halos allows
for estimates of the total stellar mass (in the central and all
satellite galaxies) --- halo mass relation (see \S
\ref{s:sat_results}).

We also make use of a secondary simulation from the Large Suite of
Dark Matter Simulations (LasDamas Project,
http://lss.phy.vanderbilt.edu/lasdamas/) in our sample variance
calculations.  The L80G simulation is too small for use in calculating
the sample variance between multiple independent mock surveys, but the
larger size of the LasDamas simulation (420 $h^{-1}$ Mpc, $1400^3$
particles) makes it ideal for this purpose.  However, the LasDamas
simulation has poorer mass resolution (a minimum particle size of
$1.9\times 10^9$ $\Msun$) and force resolution (8 $h^{-1}$ kpc),
making it unable to resolve subhalos (particularly after accretion) as
well as L80G.  The LasDamas simulation assumes a flat, $\Lambda$CDM
cosmology ($\Omega_M = 0.25$, $\Omega_\Lambda = 0.75$, $h = 0.7$, and
$\sigma_8 = 0.8$) which is very close to the WMAP5 best-fit cosmology
\citep{wmap5}.  Collisionless gravitational evolution was provided by
the \textsc{gadget}-2 code \citep{Springel05}.  Halos are identified
using friends of friends with a linking length of 0.164.  The subfind
algorithm \cite{Springel05} is used to identify substructure.

As mentioned, the primary use of the LasDamas simulation is in
sampling the halo mass functions in mock surveys to model the effects
of sample variance on high-redshift pencil-beam galaxy surveys.  The
mock surveys are constructed so as to mimic the observations in
\cite{perezgonzalez-2007}.  In each mock survey, three pencil-beam
lightcones (matching the angular sizes of the three fields in
\citealt{perezgonzalez-2007}) with random orientations are sampled from a
random origin in the simulation volume out to $z=1.3$.  Thus, by
comparing the halo mass functions in individual mock surveys to the
mass function of the ensemble, the effects of sample variance may be
calculated with full consideration of the correlations between halo
counts at different masses.

\subsubsection{Analytic Mass Functions}

\begin{figure*}
\plotsmallgrace{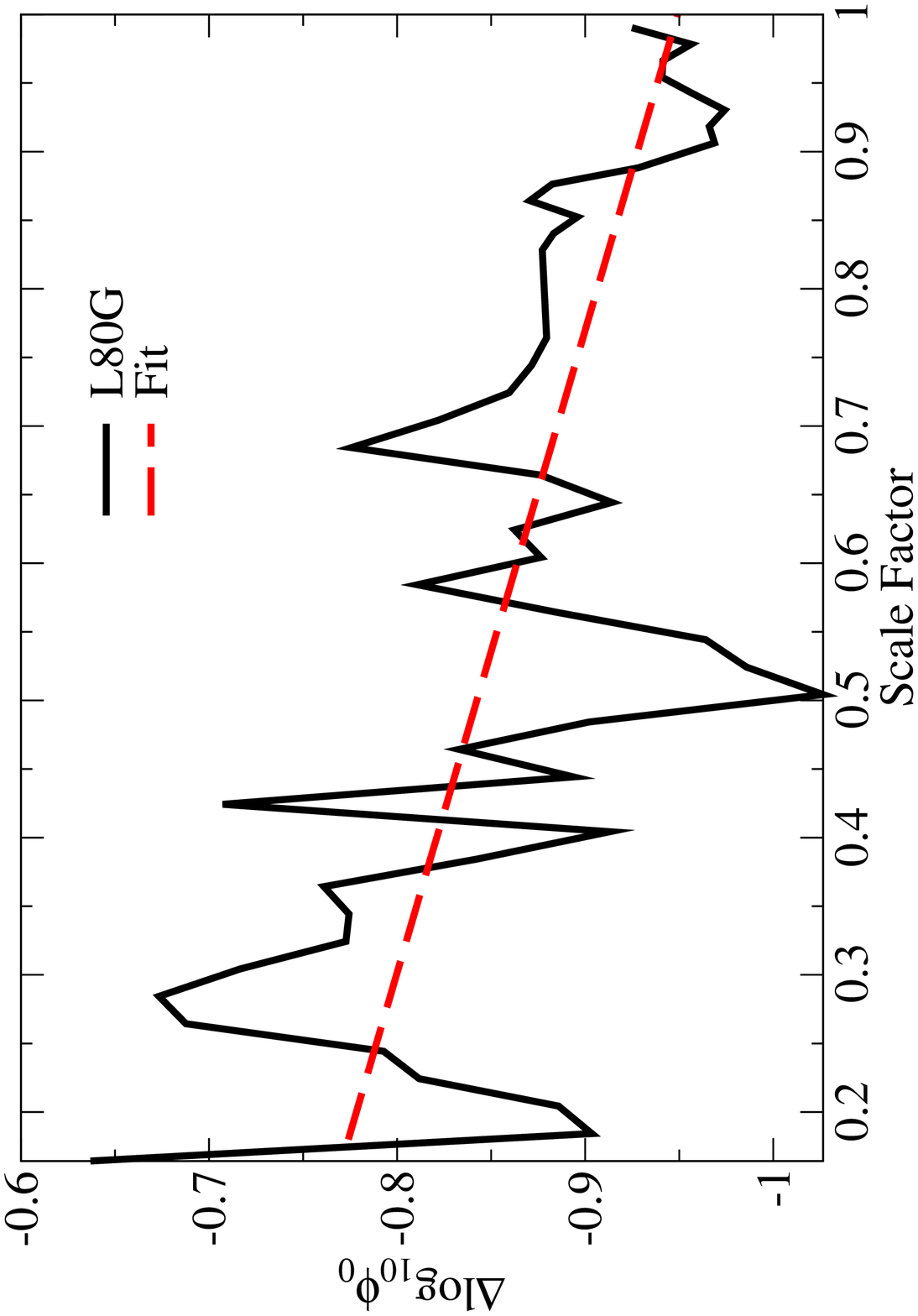}
\plotsmallgrace{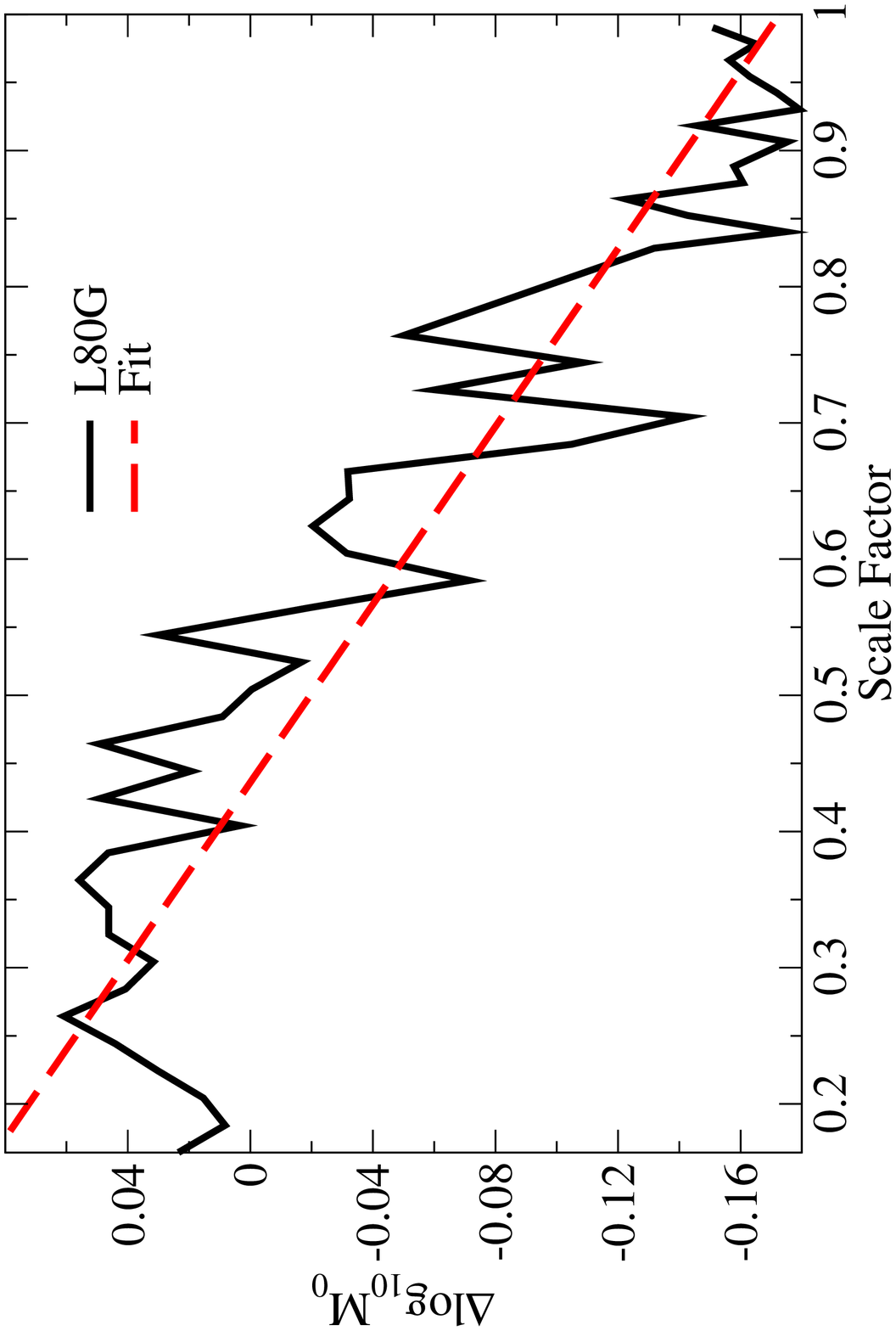}
\plotsmallgrace{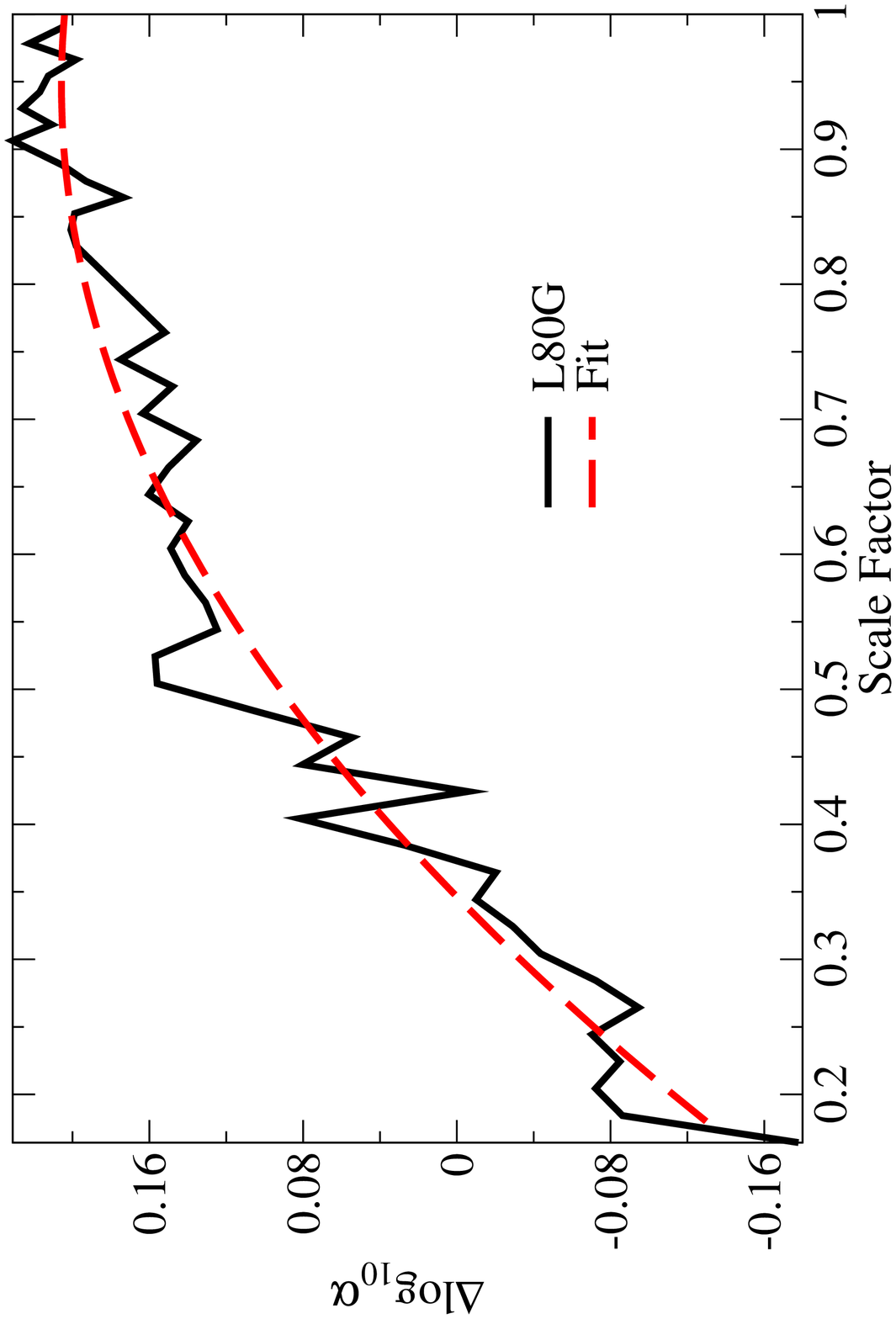}
\caption{Differences between the fitted Schechter function parameters for the satellite halo mass (at accretion) function and the central halo mass function, as a function of scale factor; e.g., $\Delta \log_{10} \phi_0$ corresponds to $\log_{10}(\phi_{0,\mathrm{sats}} / \phi_{0,\mathrm{centrals}})$. The black lines are calculated from a simulation using WMAP1 cosmology (L80G), and the red lines represent the fits to the simulation results in Equation \ref{e:sat_mf_fits}.}
\label{f:mf_fits}
\end{figure*}

\label{s:mass_functions}

The analytic mass functions of \cite{tinker-umf} are used
to calculate the abundance of halos in several cosmological models.
We calculate mass functions defined by $\mvir$, using the overdensity
specified by \cite{mvir_conv} \footnote{$\Delta_\mathrm{vir} =
  (18\pi^2 + 82 x - 39 x^2)/(1+x)$;
  $x=(1+\rho_\Lambda(z)/\rho_M(z))^{-1}-1$} This results in an
overdensity (compared to the mean background density)
$\Delta_\mathrm{vir}$ which ranges from 337 at $z=0$ to 203 at $z=1$
and smoothly approaches $180$ at very high redshifts.  Following
\cite{tinker-umf}, we use spline interpolation to calculate mass
functions for overdensities between the discrete intervals presented
in their paper.

The mass functions in \cite{tinker-umf} only include central halos.
We model the small ($\approx$20\% at $z=0$) correction to the mass
function introduced by subhalos to first order only, as the overall
uncertainty in the central halo mass function is already of order 5\%
\citep{tinker-umf}.  In particular, we calculate satellite (mass at
accretion) and central mass functions in our simulation (L80G) and fit
Schechter functions to both, excluding halos below our completeness
limit ($10^{10.3}$ $\Msun$).  Then, we plot the difference between
the Schechter parameters (the difference in characteristic mass,
$\Delta \log_{10} M_*$; the difference in characteristic density,
$\Delta \log_{10} \phi_0$; and the difference in faint-end slopes,
$\Delta \alpha$) as a function of scale factor ($a$). This gives the
satellite mass function ($\phi_s$) as a function of the central mass
function ($\phi_c$), which allows us to use this (first-order)
correction for central mass functions of different cosmologies:
\begin{equation}
\label{convert_to_sat}
\phi_s(M) = 10^{\Delta \log_{10} \phi_0} \, \left(\frac{M}{M_0 \cdot 10^{\Delta \log_{10} M_0}}\right)^{-\Delta \alpha} \, \phi_c(M / 10^{\Delta \log_{10} M_0}).
\end{equation}
From our simulation, we find fits as shown in Figure \ref{f:mf_fits}:\footnote{Comparing these fits to satellite mass functions from a more recent simulation \citep[the ``Bolshoi'' simulation]{Bolshoi}, we have verified that applying these fits to mass functions for the WMAP5 cosmology introduces errors only on the level of 5\% in overall number density, similar to the uncertainty with which the mass function is known.}
\begin{eqnarray}
\label{e:sat_mf_fits}
\Delta \log_{10} \phi_0(a) &=& -0.736 - 0.213 \, a, \nonumber\\
\Delta \log_{10} M_0(a) &=& 0.134 - 0.306 \, a,\\
\Delta \alpha(a) &=& -0.306 + 1.08 \, a - 0.570 \, a^2. \nonumber
\end{eqnarray}
The mass function used here may be 
be easily replaced by an arbitrary mass function, as
detailed in Appendix \ref{direct_matching}.

\subsubsection{Modeling Uncertainties in Cosmological Parameters}

Our fiducial results are calculated assuming WMAP5 cosmological
parameters.  In order to model uncertainties in cosmological
parameters, we have sampled an additional 100 sets of cosmological
parameters from the WMAP5+BAO+SN MCMC chains \citep[from the models
in][]{wmap5} and generated mass functions for each one according to
the method in the previous section.  Hence, to determine the variance
in the derived stellar mass -- halo mass relation caused by cosmology
uncertainties, we recalculate the relation for each sampled mass
function according to the method described in Appendix
\ref{direct_matching}.

\subsubsection{Estimating Sample Variance Effects for
the Stellar Mass Function}

\label{sample_variance}

Large--scale modes in the matter power spectrum imply that finite
surveys will obtain a biased estimate of the number densities of
galaxies and halos as compared to the full universe.  That is to say,
matching observed GSMFs measured from a finite survey to the halo mass
function estimated from a much larger volume will introduce systematic
errors into the resulting SM--HM relation.  These errors cannot be
corrected unless one has knowledge of the halo mass function for the
specific survey in question, which is in general not possible.

However, we can still calculate the uncertainties introduced by the
limited sample size.  While we cannot determine the true halo mass
function for the survey, we can calculate the probability
distribution of halo mass functions for identically shaped surveys via
sampling lightcones from simulations.  If we rematch galaxy abundances
from the observed GSMF to the abundances of halos in each of the
sampled lightcones, then the uncertainty introduced by sample variance
is exactly captured in the variance of the resulting SM--HM relations.

In detail, we create our distribution of halo mass functions by
sampling one thousand mock surveys from the LasDamas simulation (see
\S \ref{simulation}) corresponding to the exact survey parameters used
in \cite{perezgonzalez-2007}.  We fit Schechter functions to the halo
mass functions of each mock survey (over all redshifts), and we
calculate the change in Schechter parameters ($\Delta \log_{10}
\phi_0$, $\Delta \log_{10} M_0$, and $\Delta \alpha$) as compared to a
Schechter fit to the ensemble average of the mass functions.  Using
the distribution of the changes in Schechter parameters, we may mimic
to first order the expected distribution of halo mass functions for
any cosmology.  In particular, we use an equation exactly analogous to
Equation \ref{convert_to_sat} to convert the mass function for the
full universe ($\phi_\mathrm{full}$) and the distribution of $\Delta
\log_{10} \phi_0$, $\Delta \log_{10} M_0$, and $\Delta \alpha$ into a
distribution of possible survey mass functions ($\phi_{\rm obs}$):
\begin{equation}
\hspace{-3ex}\phi_{\rm obs}(M) = 10^{\Delta \log_{10} \phi_0} \, \left(\frac{M}{M_0 \cdot 10^{\Delta \log_{10} M_0}}\right)^{-\Delta \alpha} \, \phi_\mathrm{full}(M / 10^{\Delta \log_{10} M_0}).
\end{equation}

Hence, to obtain the variance in the stellar mass -- halo mass
relation caused by finite survey size, we recalculate the relation for
each one of the survey mass functions thus computed according to the
method described in Appendix \ref{direct_matching}.

\subsection{Uncertainties in Abundance Matching}
\label{s:amatch}

\subsubsection{Scatter in Stellar Mass at Fixed Halo Mass}
\label{abundance_scatter}

An important uncertainty in the abundance matching procedure is
introduced by intrinsic scatter in stellar mass at a given halo mass.
Suppose that $M_\ast(M_h)$ is the average (true) galaxy stellar mass
as a function of host halo mass.  For a perfect monotonic correlation
between stellar mass and halo mass, i.e., without scatter between
stellar and halo mass, it is straightforward to relate the true or
``intrinsic'' stellar mass function ($\ptrue$) to the halo mass
function ($\phi_h$) via
\begin{equation}
\frac{dN}{d\log_{10} M_\ast}  =  \frac{dN}{d\log_{10} M_h} \, \frac{d\log M_h}{d\log M_\ast},
\end{equation}
where $N$ is the number density of galaxies, so that
\begin{equation}
\label{chain_rule_1}
\ptrue(M_\ast(M_h)) = \phi_h(M_h) \,  \left(\frac{d\log M_\ast(M_h)}{d\log M_h}\right)^{-1}.
\end{equation}
Intuitively, as the halos of mass $M_h$ get assigned stellar masses of
$M_\ast(M_h)$, the number density of galaxies with mass $M_\ast(M_h)$
will be proportional to the number density of halos with mass $M_h$.
The above equations are simply a mathematical representation of the
traditional abundance matching technique.

Equation \ref{chain_rule_1} remains useful in the presence of scatter.
If we know the expected scatter about the mean stellar mass, say in
the form of a probability density function $P_s(\Delta \log_{10}
M_\ast | M_h)$, then we may still relate $\ptrue$ to $\phi_h$ via an
integral similar to a convolution:
\begin{eqnarray}
\ptrue(x) & = \int_{0}^\infty & \phi_h(M_h(M_\ast)) \, \frac{d \log M_h(M_\ast)}{d\log M_\ast} \times \nonumber\\
	&& \times P_s(\log_{10} \frac{x}{M_\ast} | M_h(M_\ast)) d\log_{10} M_\ast,
\end{eqnarray}
where $M_h(M_\ast)$ is the inverse function of $M_\ast(M_h)$.

\newcommand{\pdirect}{\phi_\mathrm{direct}}

This similarity to a convolution is no coincidence---mathematically,
it is analogous to how we model random statistical errors in stellar mass
measurements in \S \ref{measurement_scatter}.  Namely, if we define $\pdirect$
to equal the right-hand side of Equation \ref{chain_rule_1},
\begin{equation}
\label{pdirect}
\pdirect(M_\ast) \equiv \phi_h(M_h(M_\ast)) \, \frac{d \log M_h}{d\log M_\ast},
\end{equation}
and if we assume a probability density distribution independent of
halo mass (i.e., scatter in stellar mass at fixed halo mass is
independent of halo mass), then $\ptrue$ is exactly related to
$\pdirect$ by a convolution:
\begin{equation}
\label{no_mass_dependence}
\ptrue(M_\ast) = \int_{-\infty}^\infty \pdirect(10^y) \, P_s(y - \log_{10} M_\ast) dy,
\end{equation}
which is mathematically identical to Equation \ref{scatter_conv_eqn} in
 \S \ref{measurement_scatter}.

Then, if one calculates $\pdirect$ from $\ptrue$, one may find
$M_h(M_\ast)$ via direct abundance matching.  Namely, integrating
equation \ref{pdirect}, we have:
\begin{equation}
\label{direct_matching_eqn}
\int_{M_h(M_\ast)}^\infty \phi_h(M)d\log_{10} M = \int_{M_\ast}^\infty \pdirect(M_\ast)d\log_{10} M_\ast.
\end{equation}

Equivalently, letting $\Phi_h(M_h) \equiv \int_{M_h}^\infty \phi_h(M)
d\log_{10} M$ be the cumulative halo mass function, and letting
$\Phi_\mathrm{direct}(M_\ast) \equiv \int_{M_\ast}^\infty
\pdirect(M_\ast) d\log_{10} M_\ast$ be the cumulative ``direct''
stellar mass function, we have
\begin{equation}
\label{concise_direct_matching}
M_h(M_\ast) = \Phi_h^{-1}(\Phi_\mathrm{direct}(M_\ast)),
\end{equation}
and one may similarly find $M_\ast(M_h)$ by inverting this relation.

Our approach in all equations except for Equation
\ref{no_mass_dependence} allows a halo mass-dependent scatter in the
stellar mass, but to date the data appears to be consistent with a
constant scatter value.  For example, using the kinematics of
satellite galaxies, \cite{more-09} finds that the scatter in galaxy
luminosity at a given halo mass is $0.16\pm0.04$ dex, independent of
halo mass.  Using a catalog of galaxy groups, \cite{yang-09} find a
value of $0.17$ dex for the scatter in the stellar mass at a given
halo mass, also independent of halo mass.  Here, we thus assume a
fixed value for the scatter in stellar mass at fixed halo mass, $\xi$,
to specify the standard deviation of $P_s(\Delta \log_{10} M_\ast)$.
As the \cite{yang-09} value is consistent with the \cite{more-09}
value, we set the prior using the \cite{more-09} value and error
bounds on $\xi$, We assume a Gaussian prior on the probability
distribution for $\xi$, and we assume that the scatter itself is
log-normal.

\subsubsection{The Treatment of Satellites}
\label{s:sats}

When a galaxy is accreted into a larger system, it will likely be
stripped of dark matter much more rapidly than stellar mass because
the stars are much more tightly bound than the halo.  It has been
demonstrated that various galaxy clustering properties compare
favorably to samples of halos where satellite halos --- i.e., subhalos
--- are selected according to their halo mass at the epoch of
accretion, $M_\mathrm{acc}$, rather than their current mass
\citep[e.g.,][]{Nagai05, conroy:06, Vale06, Berrier06}.  These results
support the idea that satellite systems lose dark matter more rapidly
than stellar mass.

As commonly implemented \citep[e.g.][]{conroy:06}, the abundance
matching technique matches the stellar mass function at a particular
epoch to the halo mass function at the same epoch, using
$M_\mathrm{acc}$ rather than the present mass for subhalos.  As
$M_\mathrm{acc}$ remains fixed as long as the satellite is resolvable,
the standard technique implies that the satellite galaxy's stellar
mass will continue to evolve in the same way as for centrals of that
halo mass. Therefore, a subtle implication of the standard technique
is that satellites may continue to grow in stellar mass, even though
$M_\mathrm{acc}$ remains the same.  A different model for satellite
stellar evolution (e.g., in which stellar mass which does not evolve
after accretion) would therefore involve different choices in the
satellite matching process.

The fiducial results presented here use the standard model where
satellites are assigned stellar masses based on the current stellar
mass function and their accretion--epoch masses.  However, we also
present results for comparison in which satellite masses are assigned
utilizing the stellar mass function at the epoch of accretion,
corresponding to a situation in which satellite stellar masses do not
change after the epoch of accretion.  In order to maintain
self-consistency for the latter method, we use full merger trees (from
L80G, the simulation described in \S \ref{simulation}) to keep track
of satellites and to assure that, e.g., mergers between satellites
before they reach the central halo preserve stellar mass.

Finally, we note that any specific halo--finding algorithm may
introduce artifacts in the halo mass function in terms of when a
satellite halo is considered absorbed/destroyed.  This can have a
small effect on satellite clustering as well as number density counts.
\cite{wetzel-09} suggest an approach that avoids some of the problems
associated with resolving satellites after accretion.  Namely, they
suggest a model where satellites remain in orbit for a duration that
is a function of the satellite mass, the host mass, and the Hubble
time, after which time they dissolve or merge with the central object.
Although we have not modeled this explicitly, our satellite counts are
consistent with their recommended cutoff --- they suggest considering
a satellite halo absorbed when its present mass is less than 0.03
times its infall mass; in our simulation, only 0.1\% of all satellites
fall below this threshold.

\subsection{Functional Forms for the Stellar Mass -- Halo Mass
Relation}
\label{s:func}

In order to determine the probability distribution of our underlying
model parameters, we must first define an allowed parameter space for
the stellar mass -- halo mass relation.  Ideally, one would like a
simple, accurate, physically intuitive, and orthogonal
parameterization; in practice, we seek the best compromise with these
four goals in mind.  We consider one of the most popular methods for
choosing a functional form (indirect parameterization via the stellar
mass function) before discussing the method we use in this paper
(parameterization via deconvolution of the stellar mass function).

\subsubsection{Parameterizing the Stellar Mass Function}

In abundance matching, knowledge of the halo mass function and the
stellar mass function uniquely determines the stellar mass -- halo
mass relation.  Hence, parameterizing the stellar mass function yields
an indirect parameterization for the stellar mass -- halo mass
relation as well.  Numerous papers \citep[e.g.][]{Cole01, bell:2003,
  panter:2004, perezgonzalez-2007} have found that the GSMF is
well-approximated by a Schechter function:
\begin{equation}
\label{e:sf}
\phi(M_\ast,z) = \phi^\star(z)\left(\frac{M_\ast}{M(z)}\right)^{-\alpha(z)}\exp\left(-\frac{M_\ast}{M(z)}\right),
\end{equation}
where the Schechter parameters $\phi^\star(z)$, $M(z)$, and
$\alpha(z)$ evolve as functions of the redshift $z$.  In many previous
works on abundance matching \citep[e.g.][]{Conroy09}, it is the
Schechter function for the stellar mass function that sets the form of
the SM--HM relation.

More recently, however, several authors have noted that the GSMF
cannot be matched by a single Schechter function for $z < 0.2$ to
within statistical errors \citep[e.g.][]{li-2009, Baldry08}, in part
because of an upturn in the slope of the GSMF for galaxies below
$10^9$ $\Msun$ in stellar mass.  It is possible that a conspiracy of
systematic errors causes the observed deviations, but there is no
fundamental reason to expect the intrinsic GSMF to be fit exactly by a
Schechter function (see discussion in Appendix \ref{sample_smf_calc}).
In any case, our full parameterization ---either the stellar mass function
or the error parameterization --- \textit{must} be able to capture all
the subtleties of the observed stellar mass function.  Hence, we are
inclined to adopt a more flexible model than the Schechter function of
equation \ref{e:sf}.  Other authors, wrestling with the same problem,
have chosen to adopt multiple Schechter functions,
including the eleven-parameter triple piecewise Schechter-function fit
used by \cite{li-2009}.  While accurate, these models often add
complication without increasing intuition.

\subsubsection{Deconvolving the Observed Stellar Mass Function}

Rather than attempting to parameterize the stellar mass
function, we could use abundance matching directly to derive
the stellar mass -- halo mass relation for the maximal-likelihood
stellar mass function, and then find a fit which can parameterize the
uncertainties in the shape of the relation.
This process is complicated by the various errors which we must
take into account.  Recall from
Equations \ref{scatter_conv_eqn} and \ref{no_mass_dependence} that
\begin{equation}
\label{pobs}
\pobs(M_\ast) = \pdirect(M_\ast) \circ P_s(\Delta \log_{10} M_\ast) \circ  P(\Delta \log_{10} M_\ast),
\end{equation}
(where ``$\circ$'' denotes the convolution operation, $P_s$ is the
probability distribution for the scatter in stellar mass at fixed halo
mass, and $P$ is the probability distribution for errors in observed
stellar mass at fixed true stellar mass).  However, if we obtain
$\pdirect$ by deconvolution of the observed stellar mass function
$\pobs$, we may use direct abundance matching (Equation
\ref{concise_direct_matching}) to determine the maximum likelihood
form of $M_h(M_\ast)$.

\begin{figure}[!t]
\plotgrace{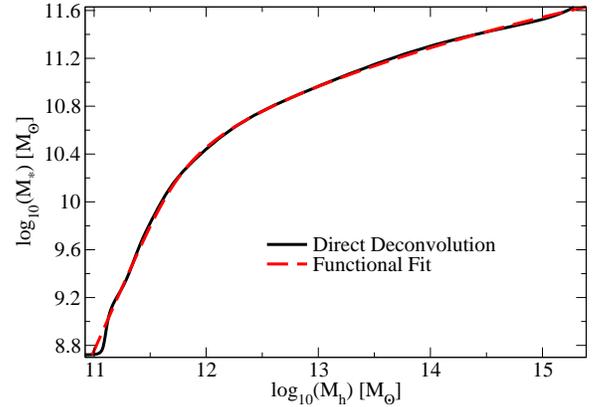}
\caption{Relation between halo mass and stellar mass in the local
  Universe, obtained via direct deconvolution of the stellar mass
  function in \cite{li-2009} matched to halos in a WMAP5 cosmology.
  The deconvolution includes the most likely value of scatter in
  stellar mass at a given halo mass as well as statistical errors in
  individual stellar masses.  The direct deconvolution ({\it solid
    line}) is compared to the best fit to Eq. \ref{eq:fit} ({\it red
    dashed line}).}
\label{fiducial_deconv}
\end{figure}

Figure \ref{fiducial_deconv} shows the result of calculating
$M_h(M_\ast)$ at $z \sim 0.1$ via deconvolution and direct matching of
the stellar mass function as described in the previous section.  We
choose the maximum-likelihood value for the distribution function
$P_s$ (namely, 0.16 dex log-normal scatter), and we use WMAP5
cosmology for the halo mass function $\phi_h$ in the derivation.

While deconvolution plus direct abundance matching gives an unbiased
calculation of the relation, there are several problems which prevent
it from being used directly to calculate uncertainties:
\begin{enumerate}
\item Deconvolution will tend to amplify statistical variations
in the stellar mass function---that is, shallow bumps in
the GSMF will be interpreted as convolutions of a sharper feature.
\item Deconvolution will give different results depending on
the boundary conditions imposed on the stellar mass function
(i.e., how the GSMF is extrapolated beyond the reported data
points)---the effects of which may be seen at the edges of
the deconvolution in Figure \ref{fiducial_deconv}.
\item Deconvolution becomes substantially more problematic
when the convolution function varies over the redshift range,
as it does for our higher-redshift data ($z>0.2$).
\item Deconvolution cannot extract the relation at a single
redshift---instead, it will only return the relation averaged
over the redshift range of galaxies in the reported GSMF.
\end{enumerate}

For these reasons, we choose to find a fitting formula instead.  In
the discussion that follows, we fit $M_h(M_\ast)$ (the halo mass for
which the average stellar mass is $M_\ast$) rather than the more
intuitive $M_\ast(M_h)$ (the average stellar mass at a halo mass
$M_h$) primarily for reasons of computational efficiency.  From
Equations \ref{pdirect} and \ref{pobs}, the calculation of what
observers would see ($\pobs$) for a trial stellar mass -- halo mass
relation requires many evaluations of $M_h(M_\ast)$ and no evaluations
of $M_\ast(M_h)$.  If we had instead parameterized $M_\ast(M_h)$, and
then inverted as necessary in the calculation of $\pobs$, our
calculations would have taken an order of magnitude more computer
time.

\subsubsection{Fitting the Deconvolved Relation}

\label{deconvolution_functions}

It is well-known from comparing the GSMF
(or the luminosity function) to the halo mass function that
high-mass ($M_\ast \gtrsim 10^{10.5}$ $\Msun$) galaxies have a
significantly different stellar mass-halo mass scaling than low-mass
galaxies, which is usually attributed to different feedback mechanisms
dominating in high-mass vs.\ low-mass galaxies.  The transition point
between low-mass and high-mass galaxies---seen as a turnover in the
plot of $M_h(M_\ast)$ around $M_\ast = 10^{10.6}$ $\Msun$ in Figure
\ref{fiducial_deconv}---defines a characteristic stellar mass
($M_{\ast,0}$) and an associated characteristic halo mass ($M_1$).
Hence, we consider functional forms which respect this general
structure of a low stellar mass regime and a high stellar mass regime
with a characteristic transition point:
\begin{eqnarray*}
 \log_{10}(M_h(M_\ast)) = & \phantom{+}\log_{10}(M_1) & \textrm{[Characteristic Halo Mass]}\\
 & + f_{low}(M_\ast / M_{\ast,0}) &\textrm{[Low-mass functional form]}\\
 & + f_{high}(M_\ast / M_{\ast,0}) \; & \textrm{[High-mass functional form]}
\end{eqnarray*}
where $f_{low}$ and $f_{high}$ are dimensionless functions dominating
below and above $M_{\ast, 0}$, respectively.

For low-mass galaxies ($M_\ast < 10^{10.5}$ $\Msun$), we find the
stellar mass -- halo mass relation to be consistent with a power--law:
\begin{eqnarray}
\frac{M_h(M_\ast)}{M_1} & \approx & \left(\frac{M_\ast}{M_{\ast,0}}\right)^\beta 
   ,\textrm{ or} \nonumber\\
\log(M_h(M_\ast)) & \approx & \log(M_1) + 
	\beta\log\left(\frac{M_\ast}{M_{\ast,0}}\right).
\end{eqnarray}

For high-mass galaxies, we find the stellar mass--halo mass relation
to be \textit{inconsistent} with a power--law.  In particular, the
logarithmic slope of $M_h(M_\ast$) changes with $M_\ast$, with $d\log
M_h / d\log M_\ast$ always increasing as $M_\ast$ increases.  This may
seem like a small detail; after all, by eye, it appears that a power
law could be a reasonable fit for high-mass galaxies in Figure
\ref{fiducial_deconv}.  In addition, because previous authors
\citep[e.g.,][]{moster-09,yang-08} have used power laws, it may not
seem necessary to use a different functional form.

In order to explore this issue, we tried a general double power--law
functional form for $M_h(M_\ast)$ which parameterized a superset of
the fits used in \cite{moster-09} and \cite{yang-08} (in particular,
the same form as in Equation \ref{double_power_law_relation} in
Appendix \ref{sample_smf_calc}).  We found that this approach had two
major problems common to any such power--law form:
\begin{enumerate}
\item As the logarithmic slope of $M_h(M_\ast)$ increases with
  increasing $M_\ast$, the best-fit power--law for high-mass galaxies
  will depend on the upper limit of $M_\ast$ in the available data for
  the GSMF.  Thus, the best-fit power--law will depend on the number
  density limit of the observational survey used---rather than on any
  fundamental physics.  Moreover, for studies such as this one which
  consider redshift evolution, the different number densities probed
  at different redshifts result in a completely artificial
  ``evolution'' of the best-fit power--law.
\item The best--fit power--law will not depend on the
  highest mass galaxies alone; instead, it will be something of an
  average over all the high-mass galaxies.  Because the logarithmic
  slope is increasing with $M_\ast$, this means that the best-fit power
  law for $M_h(M_\ast)$ will increasingly underestimate the true
  $M_h(M_\ast)$ at high $M_\ast$.  Namely, the fit will underestimate
  the halo mass corresponding to a given stellar mass, and therefore
  (as lower-mass halos have higher number densities) result in 
  stellar mass functions
  \textit{systematically} biased above observational values.  However,
  a systematic bias in our functional form will influence the best-fit
  values of the systematic error parameters.  The systematic bias
  caused by assuming a power--law form turns out to be most degenerate
  with the scatter in stellar mass at fixed halo mass ($\xi$).  As a
  result, for the MCMC chains which assumed a double power--law form
  for $M_h(M_\ast)$, the posterior distribution of $\xi$ was 0.09$\pm0.02$
  dex, which just barely lies within $2\sigma$ of the constraints from
  \cite{more-09}.
\end{enumerate}

These problems are not as significant if one only considers the
stellar mass function at a single redshift, or if one does not allow
for the systematic errors which change the overall shape of the
stellar mass function ($\kappa$, $\xi$, and $\sigma(z)$).  However, we
find that the issues listed above exclude the use of a power--law for
our purposes.  Instead, we find that $M_h(M_\ast)$ asymptotes to a
\textit{sub-exponential} function for high $M_\ast$, namely, a
function which climbs more rapidly than any power--law function, but
less rapidly than any exponential function.  We find that high--mass
galaxies ($M_\ast > 10^{10.5}$ $\Msun$) are well fit by the relation
\begin{eqnarray}
M_h(M_\ast) & \stackrel{\sim}{\propto} & 
	10^{\left(\frac{M_\ast}{M_{\ast,0}}\right)^\delta},
	\textrm{ or} \nonumber\\
\log_{10}(M_h(M_\ast)) & \to & \log_{10}(M_1) + \left(\frac{M_\ast}
	{M_{\ast,0}}\right)^\delta
\end{eqnarray}
where $\delta$ sets how rapidly the function climbs; $\delta \to 0$
would correspond to a power--law, and $\delta = 1$ would correspond to
a pure exponential.  Typical values of $\delta$ at $z=0$ range from
$0.5-0.6$.  It is not obvious what physical meaning can be directly
inferred from the choice of a sub-exponential function---after all,
the stellar mass of a galaxy is a complicated integral over the merger
and evolution history of the galaxy---but it could suggest that the
physics driving the $M_h(M_\ast)$ relation at high--mass is not
scale--free. 

Although this form now matches the asymptotic behavior for the highest
and lowest stellar mass galaxies, one additional parameter is necessary
to match the functional form of the deconvolution.
That is to say, galaxies in between the extremes in stellar mass will lie in
a transition region, as they may have been substantially affected 
by multiple feedback mechanisms.  The width of this
transition region will depend on many things---e.g., how long galaxies
take to gain stellar mass, how much of the stellar mass present came
from quiescent star formation as opposed to mergers, and the degree of
interaction between multiple feedback mechanisms.  Hence, instead of
having $M_h(M_\ast)$ become suddenly sub-exponential for galaxies
larger than $M_{\ast, 0}$, we allow for a slow ``turn-on'' of the more
rapid growth.  The behavior of $M_h(M_\ast)$ is best fit
by modifying the previous equation to 

\begin{equation}
\label{subexponential_transition}
\log_{10}(M_h(M_\ast)) \to \log_{10}(M_1) + \frac{\left(\frac{M_\ast}
	{M_{\ast,0}}\right)^\delta}{1+  \left(\frac{M_{\ast}}
		{M_\ast,0}\right)^{-\gamma}}
\end{equation}

The denominator, $1+(M_{\ast}/M_{\ast,0})^{-\gamma}$, is large for
$M_\ast < M_{\ast,0}$, and it falls to unity for $M_\ast >
M_{\ast,0}$ at a rate controlled by $\gamma$.  A larger value of
$\gamma$ implies a more rapid transition between the power--law
and sub-exponential behavior (typical values for ($\gamma$)
at $z=0$ are 1.3-1.7).  As the non-constant piece of
$M_h(M_\ast)$ in Equation \ref{subexponential_transition} is
$\frac{1}{2}$ for $M_\ast = M_{\ast,0}$, we add a final
factor of $-\frac{1}{2}$ to compensate so that 
$M_h(M_{\ast,0}) = M_1$.

To summarize, our resulting best--fit functional form has five
parameters:
\begin{displaymath}
 \log_{10}(M_h(M_\ast)) = \hspace{0.65\columnwidth}
 \end{displaymath}
 \vspace{-3ex}
\begin{equation}
 \label{functional_form}
\quad \log_{10}(M_1) + \beta\,\log_{10}\left(\frac{M_\ast}{M_{\ast,0}}\right) +
 \frac{\left(\frac{M_\ast}{M_{\ast,0}}\right)^\delta}{1 + \left(\frac{M_{\ast}}{M_{\ast,0}}\right)^{-\gamma}} - \frac{1}{2}.
\label{eq:fit}
\end{equation}
Where $M_1$ is a characteristic halo mass, $M_{\ast,0}$ is a
characteristic stellar mass, $\beta$ is the faint-end slope, and
$\delta$ and $\gamma$ control the massive-end slope.  The best fit
using this functional form is shown in Figure \ref{fiducial_deconv},
and it achieves excellent agreement over the entire range of stellar
masses.

Deconvolving the GSMF at higher redshifts does not suggest that
anything more than linear evolution in the parameters is necessary, at
least out to $z=1$.  While the characteristic mass of the GSMF and the
characteristic mass of the halo mass function certainly evolve, the
change in the \textit{shapes} of the two functions is relatively
slight.  As we wish for the functional form to have a natural
extension to higher redshifts, we parameterize the evolution in terms
of the scale factor ($a$):
\begin{eqnarray}
\log_{10}(M_1(a)) & = & M_{1,0} + M_{1,a} \, (a-1), \nonumber\\
\log_{10}(M_{\ast,0}(a)) & = & M_{\ast,0,0} + M_{\ast,0,a} \, (a-1), \nonumber \\
\beta(a) & = & \beta_0 + \beta_a \, (a-1),\\ 
\delta(a) & = & \delta_0 + \delta_a \, (a-1), \nonumber \\
\gamma(a) & = & \gamma_0 + \gamma_a \, (a-1),  \nonumber
\label{eq:zscaling}
\end{eqnarray}
where $a=1$ is the scale factor today.

\begin{deluxetable*}{rccc}
\tablecaption{Summary of Model Parameters
\label{parameters_table}}
\tablehead{Symbol & Description & Prior\tablenotemark{a} & Section}
$M_h(M_\ast)$ & The halo mass for which the average 
stellar mass is $M_\ast$ & N/A & \ref{deconvolution_functions}\\
\hline
$M_1$ & Characteristic Halo Mass & Flat (Log) & \ref{deconvolution_functions}\\
$M_{\ast,0}$ & Characteristic Stellar Mass & Flat (Log) & \ref{deconvolution_functions}\\
$\beta$ & Faint-end power law ($M_h \sim M_\ast^\beta$) & Flat (Linear) & \ref{deconvolution_functions}\\
$\delta$ & Massive-end sub-exponential ($\log_{10}(M_h) \sim M_\ast^\delta$) & Flat (Linear) & \ref{deconvolution_functions}\\
$\gamma$ & Transition width between faint- and massive-end relations & Flat (Linear) & \ref{deconvolution_functions}\\
$(x)_0$ & Value of the variable $(x)$ at the present epoch, where $(x)$ is one of $(M_1, M_{\ast,0}, \beta, \delta, \gamma)$ & (see
above) &\ref{deconvolution_functions}\\
$(x)_a$ & Evolution of the variable $(x)$ with scale factor & (same as for $(x)_0$) &\ref{deconvolution_functions}\\
\hline
$\mu$ & Systematic offset in $M_\ast$ calculations & $G(0,0.25)$ (Log) & \ref{systematics_sm}\\
$\kappa$ & Systematic mass-dependent offset in $M_\ast$ calculations & $G(0,0.10)$ (Linear) & \ref{systematics_sm}\\
$\sigma_z$ & Redshift scaling of statistical errors in $M_\ast$ calculations & $G(0.05, 0.015)$ (Log) & \ref{measurement_scatter}\\
$\xi$ & Scatter in $M_\ast$ at fixed $M_\mathrm{h}$ &$G(0.16, 0.04)$ (Log) & \ref{abundance_scatter}
\tablenotetext{a}{See Equations \ref{eq:systematics},
  \ref{sigma_evolution}, \ref{eq:fit}-\ref{eq:zscaling}.  $G(x,s)$
  denotes a Gaussian prior centered at $x$ with standard deviation
  $s$, in either linear or logarithmic units.  `Flat' denotes a uniform prior in either linear or logarithmic
  units.}
\end{deluxetable*}

\subsection{Calculating Model Likelihoods}
\label{s:mcmc}

We make use of a Markov Chain Monte Carlo (MCMC) method to generate a
probability distribution in our complete parameter space of stellar
mass function parameters $(M_{1,0}, M_{1,a}, M_{\ast,0,0},
M_{\ast,0,a}, \beta_0, \beta_a, \delta_0, \delta_a, \gamma_0,
\gamma_a)$, systematic modeling errors $(\kappa, \mu, \sigma_z)$, and
the scatter in stellar mass at fixed halo mass $(\xi)$.  A brief
summary of each of these parameters appears in Table
\ref{parameters_table} along with a reference to the section in which
it was first described.  Using this full model, we may calculate the
stellar mass functions expected to be seen by observers
($\phi_\mathrm{expect}$) for a large number of points in parameter
space, and compare them to observed GSMFs
\citep{li-2009,perezgonzalez-2007}.  Note that, as the observational
data always covers a range of redshifts, we must mimic this in our
calculation of $\phi_\mathrm{expect}$:
\begin{equation}
\phi_\mathrm{expect} = \frac{\int_{z_1}^{z_2} \phi_\mathrm{fit}(z) dV_C(z)}{\int_{z_1}^{z_2} dV_C(z)},
\end{equation}
where $dV_C(z)$ is the comoving volume element per unit solid angle
as a function of redshift.  Then, we can write the likelihood as 
$\mathcal{L} = \exp\left(-\chi^2/2\right)$, where
\begin{equation}
\chi^2  = \int \left[\frac{\log_{10}[\phi_\mathrm{expect}(M_\ast) / \pobs(M_\ast)]}{\sigma_\mathrm{obs}(M_\ast)}\right]^2 d\log_{10}(M_\ast), 
\label{chi2}
\end{equation}
and where $\sigma_\mathrm{obs}(M_\ast)$ is the reported statistical error in
$\pobs$ as a function of stellar mass.

Note that, as defined above, the equation for $\chi^2$ contains the
assumption that there is only one independent observation point for
the GSMF per decade in stellar mass (from the weight of
$d\log_{10}(M)$).  We may tune this assumption introducing another
parameter $n$---the number of non-correlated observations per decade
in stellar mass---which would change the likelihood function to
$\mathcal{L} = \exp\left(-n\chi^2/2\right)$.  Here, we assume that
each of the data points reported by \cite{li-2009} and
\cite{perezgonzalez-2007} are independent---such that $n=10$ for the
former paper and $n=5$ for the latter paper.

The MCMC chains each contain $2^{22} \approx 4\times 10^6$ points.
We verify convergence according to the algorithm in \cite{dunkley-2005};
in all cases, the ratio of the sample mean variance to the distribution
variance (the ``convergence ratio'') is below 0.005.

\subsection{Methodology Summary}

\label{s:summary}

Our procedure to calculate the stellar mass -- halo mass relation,
taking into account all mentioned uncertainties, may be summarized in
seven steps:

\begin{enumerate}
\item \label{trial_point} We select a trial point in the parameter
  space of SM--HM relations as well as a trial point in our parameter
  space of systematics ($\mu$, $\kappa$, $\sigma_z$, $\xi$).  A
  complete list of parameters and descriptions is given in Table
  \ref{parameters_table}.
\item \label{direct_point}
 The trial SM--HM relation gives a one-to-one mapping between halo
  masses and stellar masses, giving a direct conversion from the halo
  mass function to a trial galaxy stellar mass function (corresponding
  to $\pdirect$ in \S \ref{abundance_scatter}).
\item This trial GSMF is convolved with the probability distributions
  for scatter in stellar mass at fixed halo mass (controlled by $\xi$,
  see \S \ref{abundance_scatter}) and for scatter in
  observer-determined stellar mass at fixed true stellar mass
  (partially controlled by $\sigma_z$, see \S
  \ref{measurement_scatter}).
\item \label{systematics_point}
   The resulting GSMF is shifted by a uniform offset in stellar
  masses (controlled by $\mu$) to account for uniform systematic
  differences between our adopted stellar masses and the true
  underlying masses.  Also, its shape is stretched or compressed to
  account for stellar mass--dependent offsets between our masses and
  the true underlying masses (controlled by $\kappa$, see \S
  \ref{systematics_sm}).
\item We repeat steps \ref{direct_point}-\ref{systematics_point} for
  all redshifts in the range covered by the observed data set.  We may
  then calculate the expected GSMF in each redshift bin for which
  observers have reported data.  The likelihood of the expected GSMFs
  given the measured GSMFs is then used to determine the next step in
  the MCMC chain.
\item To account for sample variance in the observed stellar mass
  functions above $z \sim 0.2$, we recalculate each SM--HM relation 
  in the chain for an alternate halo mass function taken from a randomly
  sampled mock survey (see \S \ref{sample_variance}) and re-fit our
  functional form to the redshift evolution of the relation.
  Similarly, for the results
  which include cosmology uncertainty, we recalculate each SM--HM
  relation for an alternate halo mass function randomly selected from
  the MCMC chain used to determine the WMAP5 cosmology uncertainties.
\label{mcmc_point}
\item We repeat steps \ref{trial_point}-\ref{mcmc_point} to build a
  joint probability distribution for the SM--HM relation and the
  systematics parameter space.  The steps are repeated until the joint
  probability distribution has converged to the underlying posterior
  distribution.
\end{enumerate}

\section{Results for $0<\lowercase{z}<1$}
\label{s:results}

We now present the results of this approach to determine the SM--HM
relation and related quantities.  In \S \ref{s:gsmf}, we compare GSMFs
generated from our best fits to observed data and comment on the
effects of systematic observational biases.  We present our
best-fitting results for the SM--HM relation with full error bars in
\S \ref{smmr_results}.  We evaluate the relative importance of each of
the contributing types of error in \S \ref{s:uncertainties_impact} and
summarize the most relevant contributions in \S
\ref{s:uncertainties_summary}.  Finally, our derived SM--HM relation
is compared to other published results in \S \ref{s:comparison}.

\subsection{Galaxy Stellar Mass Functions}
\label{s:gsmf}

\begin{figure}[!t]
\plotgrace{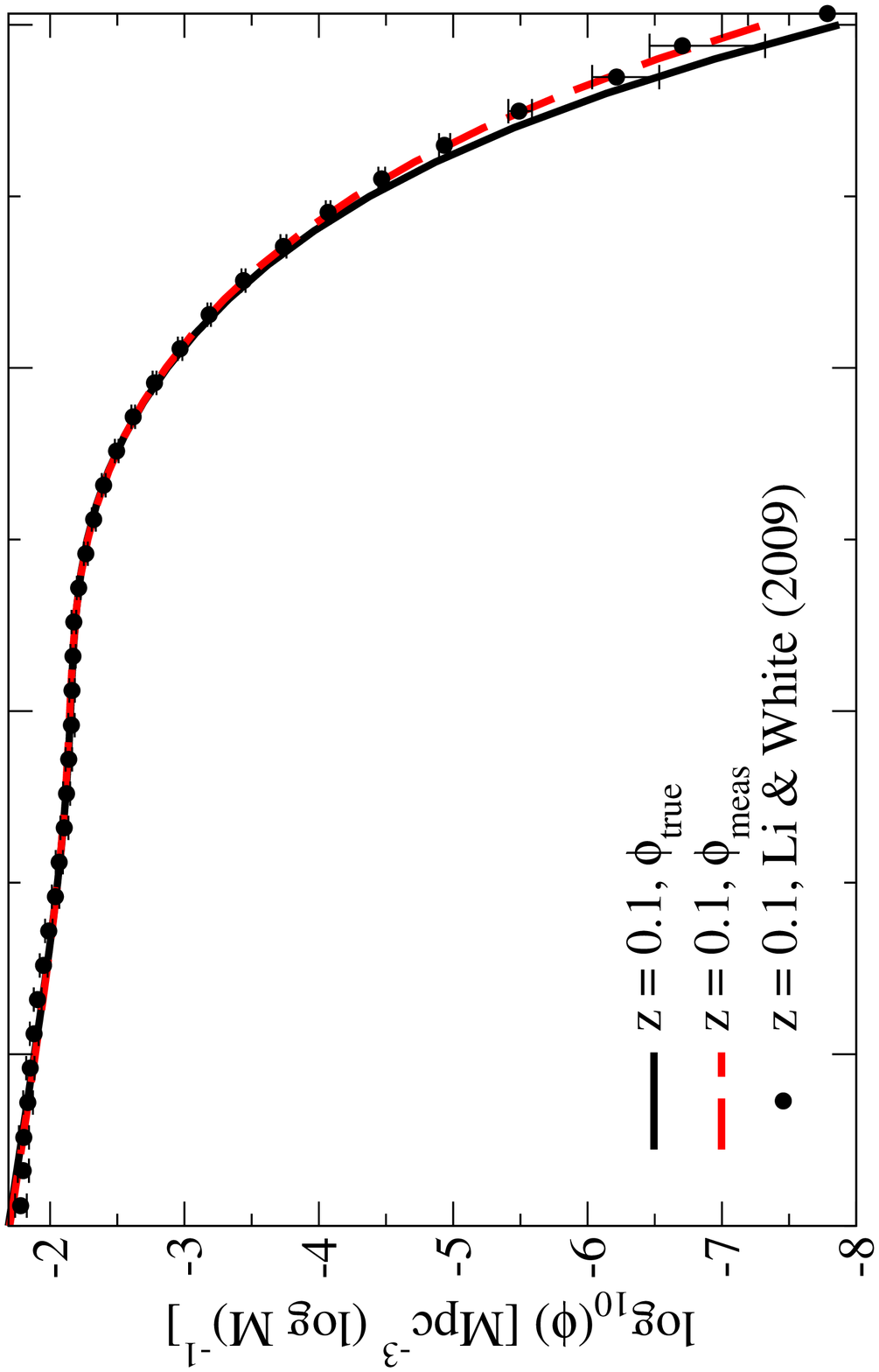}\\[-7ex]
\plotgraceflip{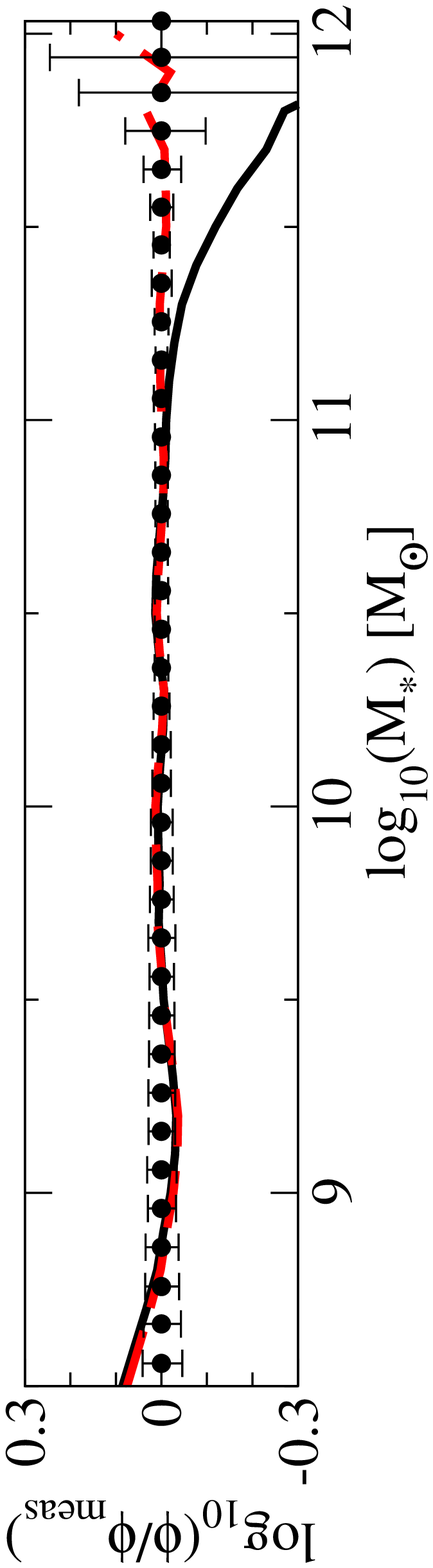}
\caption{Comparison of the best fit $\ptrue$ (the true or
  ``intrinsic'' GSMF) in our model to the resulting $\pobs$ (what an
  observer would report for the GSMF, which includes the effects of
  the systematic biases $\mu, \kappa$, and $\sigma$) at $z=0$.  Since
  the best--fit values of $\mu$ and $\kappa$ are very close to zero,
  the difference between $\pobs$ and $\ptrue$ almost exclusively comes
  from the uncertainty in measuring stellar masses ($\sigma$).}
\label{smf_sigma_comp}
\vspace{0.5cm}
\end{figure}

\begin{figure}[t!]
\plotgrace{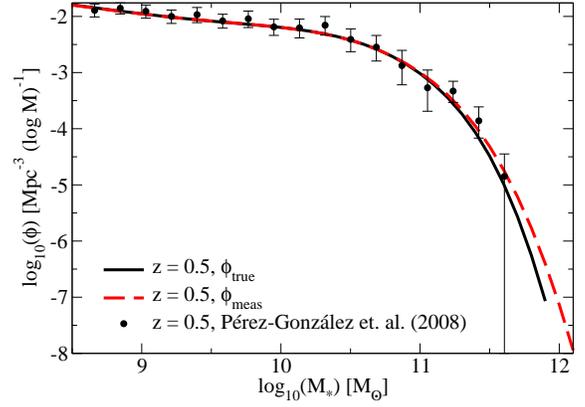}\\
\plotgrace{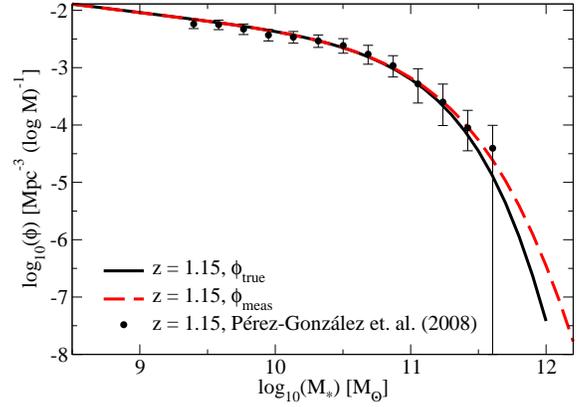}\\
\caption{Comparison of the best fit $\ptrue$ (the true or
  ``intrinsic'' GSMF) to the resulting $\pobs$ (as in Figure
  \ref{smf_sigma_comp}), for $z=0.5$ and $z=1.15$. Statistical errors
  in individual stellar masses have a larger effect at higher redshift,
  resulting in a steeper intrinsic bright end than measured.}
\label{smf_sigma_comp2}
\end{figure}

To demonstrate that our functional form for $M_h(M_\ast)$ is capable
of reproducing observed galaxy stellar mass functions, we show a
comparison between our best--fit models and the observed data in
Figs.\ \ref{smf_sigma_comp} and \ref{smf_sigma_comp2} at several
redshifts.  For our best-fit models, both $\ptrue$ (the true or
``intrinsic'' stellar mass function) and $\pobs$ (the GSMF that
observers would measure) are shown.  Recall that $\pobs$ incorporates
the effects of the systematic observational biases; namely, the
overall shift in stellar mass calculations, $\mu$, the linearly
mass-dependent shift, $\kappa$, and the statistical errors in stellar mass
calculations for individual galaxies, $\sigma(z)$.  The fact that the
best-values of the systematic parameters ($\mu$, $\kappa$, $\xi$,
$\sigma_z$) are very close to the centers of their prior distributions
provides confirmation that the functional form for the SM--HM
relation outlined in \S \ref{deconvolution_functions} does not bias
our best-fit results.

As our best-fit values for $\mu$ and $\kappa$ are close to zero (see
Table \ref{smf_ev_fit_table}), the difference between $\ptrue$ and
$\pobs$ is almost exclusively due to the scatter $\sigma(z)$ in
calculated stellar masses.  The difference between $\ptrue$ and
$\pobs$ only becomes evident for galaxies above $10^{11}$ $\Msun$,
where the falling slope of the GSMF becomes severe enough for the
scatter $\sigma(z)$ to significantly raise number counts in the
observed GSMF.  At $z\sim 0$, the systematic effect of $\sigma(z)$
puts the intrinsic GSMF well below the small statistical error bars.

At higher redshifts, although the effect of $\sigma(z)$ is larger,
current surveys at $z>0.2$ do not yet cover sufficient volume to
constrain the shape of the GSMF well at the massive end.  Nonetheless,
for future wide-field surveys at $z>0.2$, correction to the GSMF for
scatter in calculated stellar masses will be an important
consideration.

\begin{figure}[t!]
\plotgrace{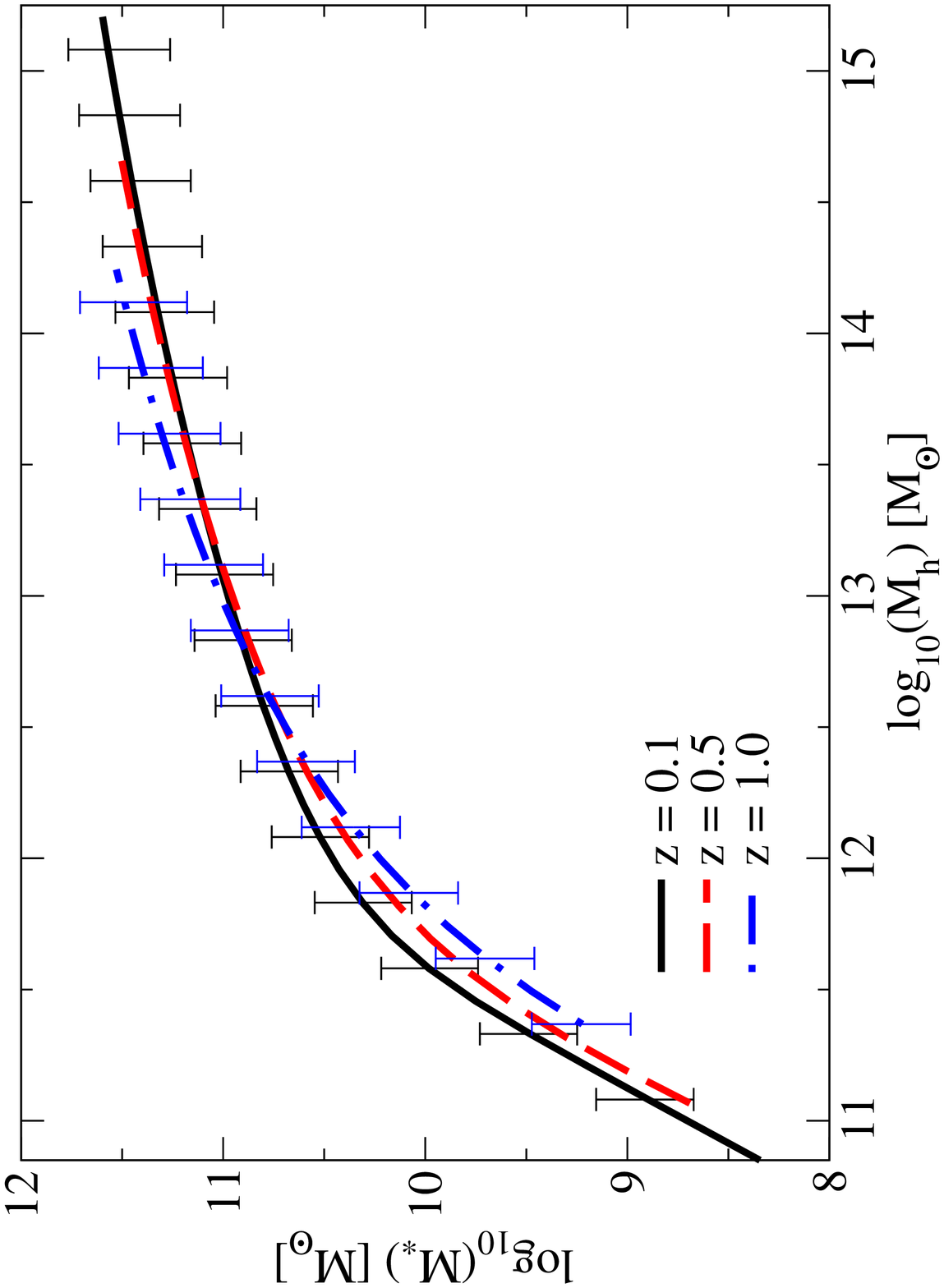}\\
\plotgrace{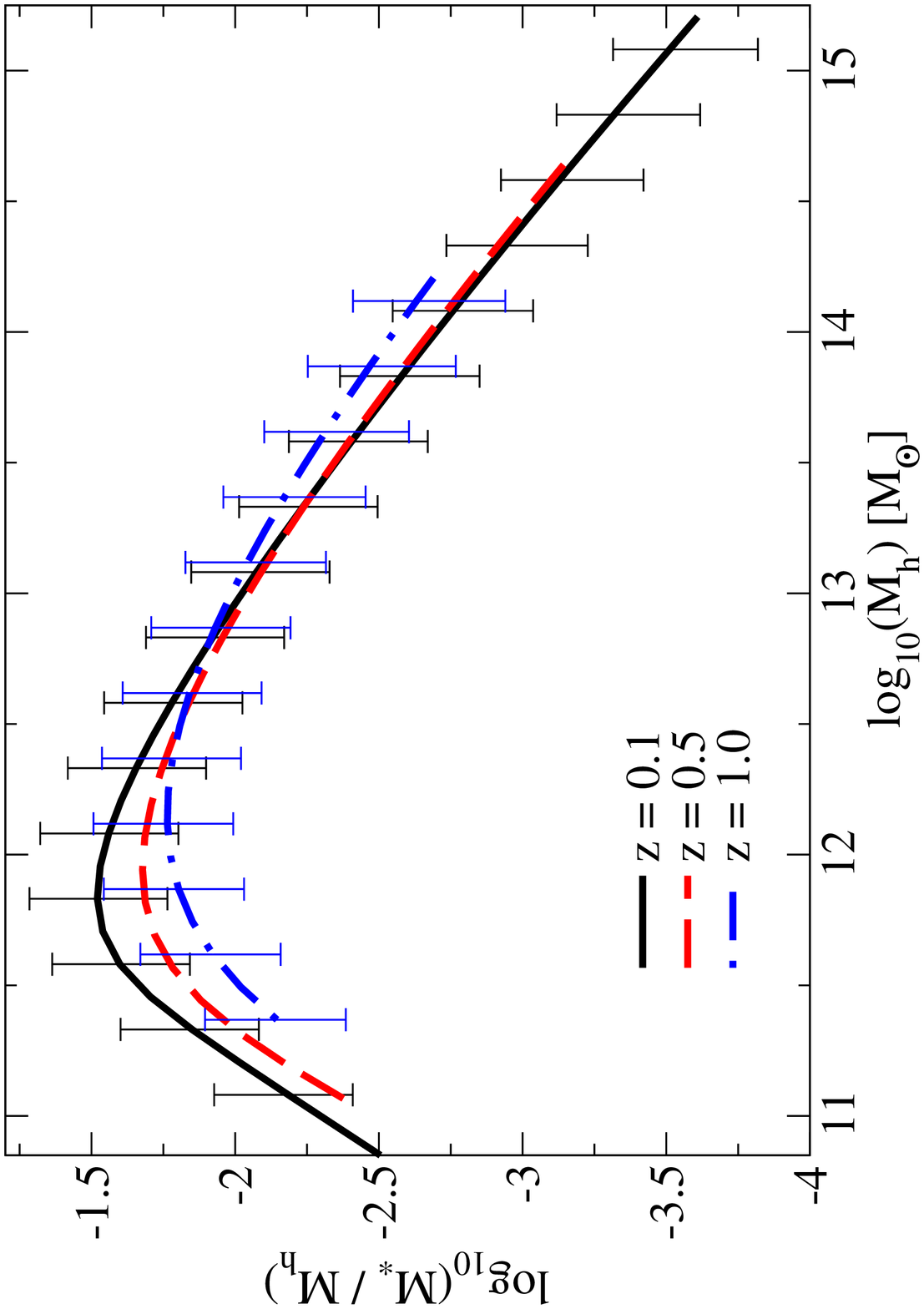}
\caption{{\it Top panel}: Stellar mass -- halo mass relation as a
  function of redshift for our preferred model.  {\it Bottom panel}:
  Evolution of the derived stellar mass fractions ($M_\ast / M_h$).
  In each case, the lines show the mean values for central galaxies.
  These relations also characterize the satellite galaxy population if
  the horizontal axis is interpreted as the halo mass at the time of
  accretion.  Errors bars include both systematic and statistical
  uncertainties, calculated for a fixed cosmological model (with WMAP5
  parameters).  }
\label{smmr_together}
\end{figure}

\begin{figure}[t!]
\plotgrace{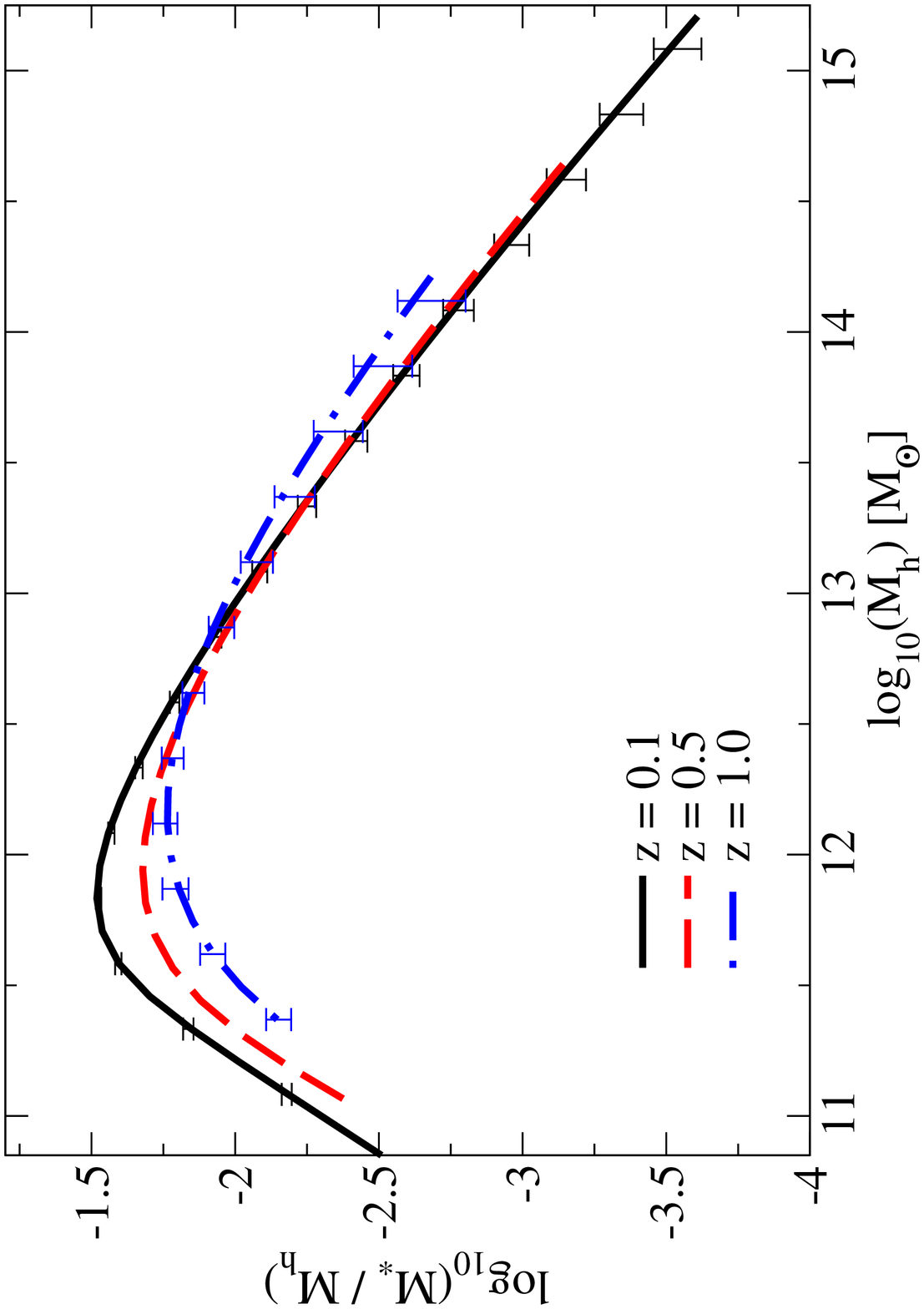}
\caption{ Evolution of the derived stellar mass fractions ($M_\ast /
  M_h$) in the absence of systematic errors.  This result is analogous
  to Figure \ref{smmr_together}, bottom panel, calculated under the
  assumption that the true values of the systematics $\mu$ and
  $\kappa$ in the stellar mass function are zero at all redshifts.}
\label{smmr_together_n}
\end{figure}

\subsection{The Best-Fit Stellar Mass -- Halo Mass Relations}
\label{smmr_results}

We plot the average stellar mass as a function of halo mass for
$z=0-1$ in Figure \ref{smmr_together} to show the evolution of the
stellar mass -- halo mass relation.  Note that as the stellar mass at
a given halo mass has a log-normal scatter (see
$\S$\ref{abund_match_uncertainties}), we use geometric averages for
stellar masses rather than linear ones.  To highlight the effects of
halo mass on star formation efficiency, we also present the SM--HM
relation in terms of the average stellar mass fraction (stellar mass /
halo mass) for $z=0-1$ as a function of halo mass in the same figure.
We focus on this quantity for the remainder of the paper.  The
best-fit parameters for the function $M_h(M_\ast)$ are given in Table
\ref{smf_ev_fit_table}, and the numerical values for the stellar mass
fractions are listed in Appendix \ref{a_data}.

\begin{figure*}[t!]
\plotgrace{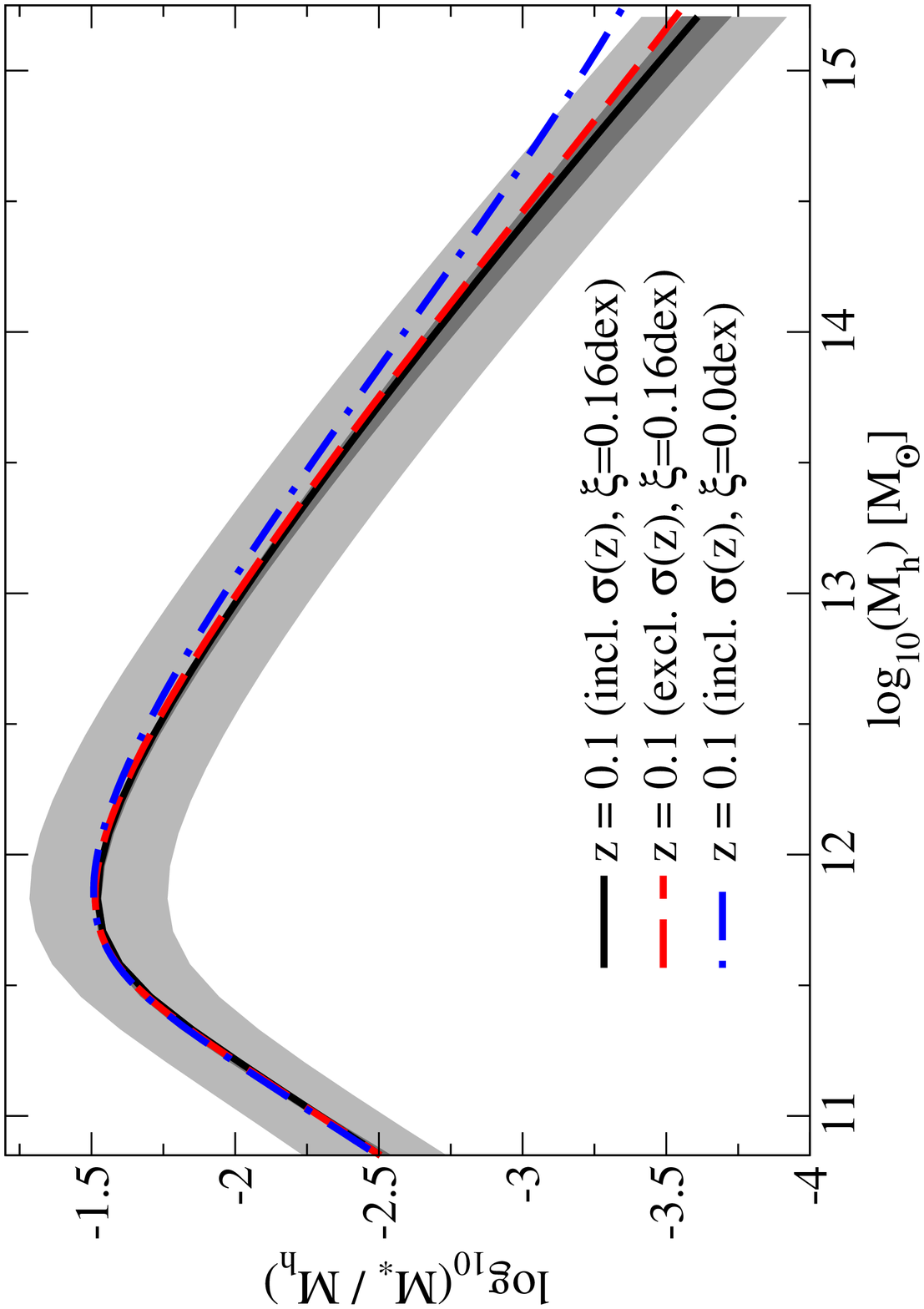}
\plotgrace{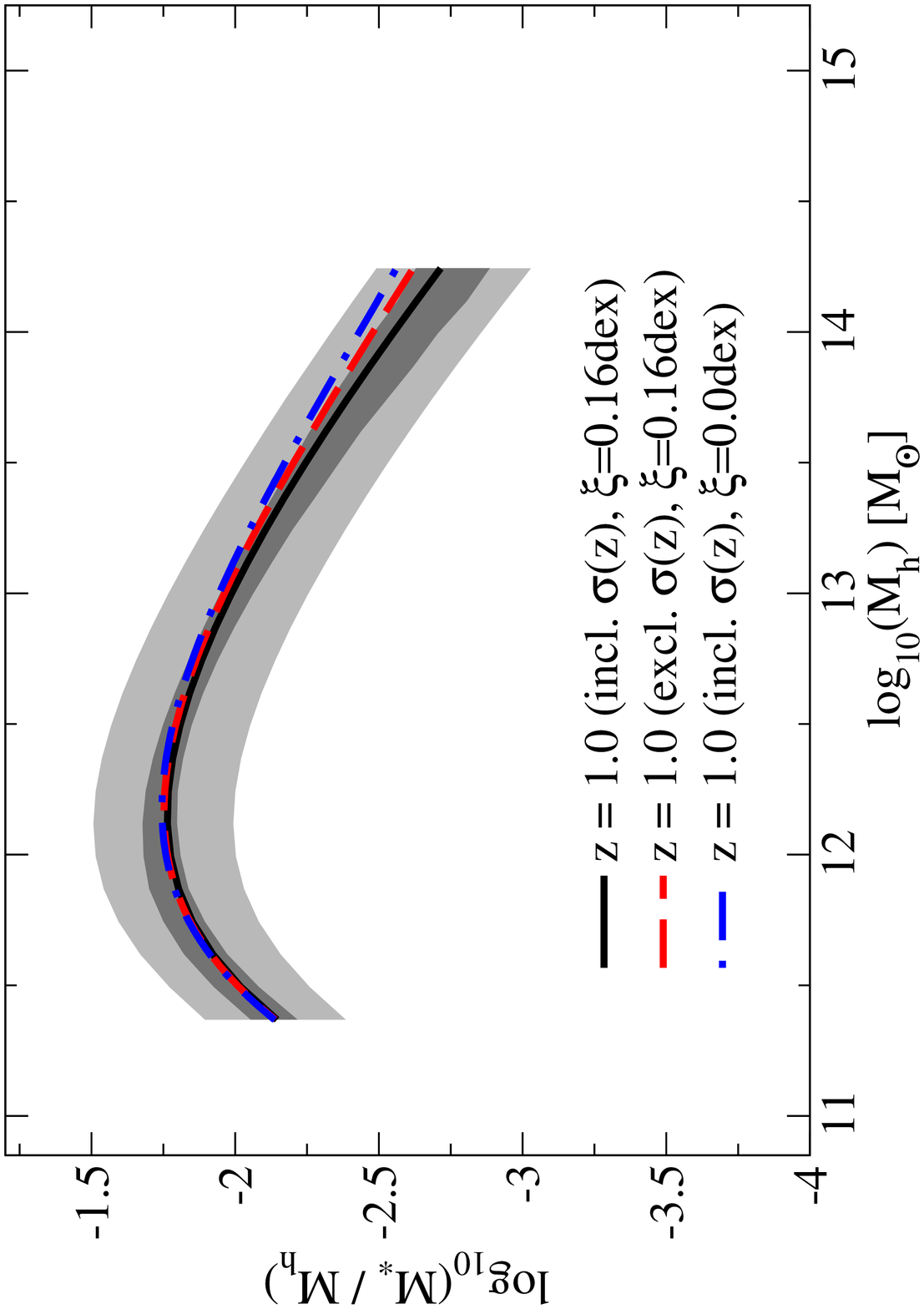}
\caption{Comparison between SM--HM relations derived in the preferred
  model (including the effects of the statistical errors $\sigma(z)$ and taking the
  scatter in stellar mass at a given halo mass to be $\xi =
  0.16$dex) to those excluding the effects of $\sigma(z)$ or taking
  $\xi = 0$, at $z=0$ ({\it left panel}) and $z=1$ ({\it right
    panel}).  Light shaded regions denote 1-$\sigma$ errors including
  both systematic and statistical errors; dark shaded regions denote
  the 1-$\sigma$ errors if the systematic offsets in stellar mass
  calculations ($\mu$ and $\kappa$) are fixed to 0.}
\label{smmr_sigma_comp}
\end{figure*}

\begin{table}
\begin{center}
\caption{Best fits for the redshift evolution of $M_h(M_\ast)$}
\label{smf_ev_fit_table}
\begin{tabular}{rccc}
\hline
\hline
Parameter & Free $(\mu, \kappa)$ & $\mu = \kappa = 0$ & Free $(\mu, \kappa)$\\
& $0<z<1$ & $0<z<1$ & $0.8 < z < 4$\\
\hline
$M_{\ast,0,0}$ & $10.72^{+0.22}_{-0.29}$ & $10.72^{+0.02}_{-0.12}$ & $11.09^{+0.54}_{-0.31}$\\
$M_{\ast,0,a}$ & $0.55^{+0.18}_{-0.79}$ & $0.59^{+0.15}_{-0.85}$ & $0.56^{+0.89}_{-0.44}$\\
$M_{\ast,0,a^2}$ & \textsc{n/a} & \textsc{n/a} & $6.99^{+2.69}_{-3.51}$\\
$M_{1,0}$ & $12.35^{+0.07}_{-0.16}$ & $12.35^{+0.02}_{-0.15}$ & $12.27^{+0.59}_{-0.27}$\\
$M_{1,a}$ & $0.28^{+0.19}_{-0.97}$ & $0.30^{+0.14}_{-1.02}$ & $-0.84^{+0.87}_{-0.58}$\\
$\beta_0$ & $0.44^{+0.04}_{-0.06}$ & $0.43^{+0.01}_{-0.05}$ & $0.65^{+0.26}_{-0.20}$\\
$\beta_a$ & $0.18^{+0.08}_{-0.34}$ & $0.18^{+0.06}_{-0.34}$ & $0.31^{+0.38}_{-0.47}$\\
$\delta_0$ & $0.57^{+0.15}_{-0.06}$ & $0.56^{+0.14}_{-0.05}$ & $0.56^{+1.33}_{-0.29}$\\
$\delta_a$ & $0.17^{+0.42}_{-0.41}$ & $0.18^{+0.41}_{-0.42}$ & $-0.12^{+0.76}_{-0.50}$\\
$\gamma_0$ & $1.56^{+0.12}_{-0.38}$ & $1.54^{+0.03}_{-0.40}$ & $1.12^{+7.47}_{-0.36}$\\
$\gamma_a$ & $2.51^{+0.15}_{-1.83}$ & $2.52^{+0.03}_{-1.89}$ & $-0.53^{+7.87}_{-2.50}$\\
\hline
$\mu$ & $0.00^{+0.24}_{-0.25}$ & \textsc{n/a} & $0.00^{+0.25}_{-0.25}$\\
$\kappa$ & $0.02^{+0.11}_{-0.07}$ & \textsc{n/a} & $0.00^{+0.14}_{-0.04}$\\
$\xi$ & $0.15^{+0.04}_{-0.02}$ & $0.15^{+0.04}_{-0.01}$ & $0.16^{+0.07}_{-0.01}$\\
$\sigma_z$ & $0.05^{+0.02}_{-0.01}$ & $0.05^{+0.02}_{-0.01}$ & $0.05^{+0.02}_{-0.01}$\\

\hline
\end{tabular}
\end{center}
\tablecomments{See Table \ref{parameters_table} and Equations \ref{eq:systematics}, \ref{sigma_evolution}, \ref{eq:fit}-\ref{eq:zscaling}, \ref{eq:fit2}.}
\end{table}

The stellar mass fractions for central galaxies consistently show a
maximum for halo masses near $10^{12}$ $\Msun$.  While the location of
this maximum evolves with time, it clearly illustrates that
star-formation efficiency must fall off for both higher and lower-mass
halos. The slopes of the SM--HM relation above and below this
characteristic halo mass are indicative of at least two processes
limiting star-formation efficiency, although mergers complicate direct
analysis for high-mass halos.  At the low-mass end, the SM--HM
relation scales as $M_\ast \sim M_h^{2.3}$ at $z=0$ and as $M_\ast
\sim M_h^{2.9}$ at $z=1$.  However, given the lack of information
about low stellar-mass galaxies at $z>0.5$, the statistical
significance of this evolution is weak; no evolution in the low-mass
slope of the relation is consistent within our one-sigma errors.
Several studies (most recently, \citealt{Baldry08,Drory09}) have
reported that the GSMF has an upturn in slope for very low stellar
masses, particularly below $10^{8.5} \Msun$; this would imply that our
best fits may overestimate the scaling relation for galaxies below
$10^{8.5} \Msun$.  At the high mass end, our best fitting function
results in a progressively shallower relation for the growth of
stellar mass with halo mass, so that no single power law can describe
the scaling.  However, for halos close to $10^{14}\Msun$, the best-fit
relation scales locally as $M_\ast \sim M_h^{0.28}$ at $z=0$ and
$M_\ast \sim M_h^{0.34}$ at $z=1$, in accord with previous studies
(see \S \ref{s:comparison}).  The results for high mass halos
are also consistent with no evolution in the 
slope of the SM--HM relation.

Figure \ref{smmr_together_n} shows the stellar mass fraction for $0 <
z < 1$ excluding the effects of systematic shifts in stellar mass
calculations (i.e., assuming $\mu = \kappa =0$).  Under the assumption
that systematic errors in stellar mass calculations result in similar
biases in stellar masses at $z=0$ as they do at higher redshifts, this
allows us to consider the evolution in normalization of the SM--HM
relation.  Low-mass halos (below $10^{12}$ M$_\odot $) display clearly
higher stellar mass fractions at late redshifts than they do at early
redshifts.  By contrast, the evolution in stellar mass fractions for
high mass halos (above $10^{13.5}$ M$_\odot $) is not statistically
significant, and it is constrained to be substantially less than for
low-mass halos.  In the time since $z=1$, this means that the star
formation rates for high-mass halos typically fall relative to
their dark matter accretion rates, whereas the opposite is true for
low-mass halos \citep{cw-08}.  The best-fitting parameters for the
SM--HM relation assuming $\mu = \kappa = 0$ appear in Table
\ref{smf_ev_fit_table}, and the data points in Figure
\ref{smmr_together_n} appear in Appendix \ref{a_data}.

\subsection{Impact of Uncertainties}
\label{s:uncertainties_impact}
\subsubsection{Systematic Shifts in Stellar Mass Calculations}

By far the largest contributor to the error budget of the SM--HM
relation is the systematic error parameter $\mu$. As the effect of
$\mu$ is to multiply all stellar masses by a constant factor, and as
the width of the error bars in Figure \ref{smmr_together} corresponds
almost exactly to the prior on $\mu$, we may conclude that reducing
the error on the systematic shifts in stellar mass calculations would
represent the single largest improvement in our understanding of the
shape of the SM--HM relation.  Figure \ref{smmr_together_n} shows
the substantially smaller error bars that result if systematic errors
($\mu$ and $\kappa$) in the stellar mass calculations are neglected.

\subsubsection{Scatter in Stellar Mass at Fixed Halo Mass}

The effect of ignoring scatter in stellar mass at fixed halo mass
(i.e., setting $\xi=0$) is shown at two redshifts in Figure
\ref{smmr_sigma_comp}.  We find that the change is insignificant below
halo masses of $10^{12}$ $\Msun$, and is within statistical error
bars below $10^{13}$ $\Msun$ for $z=1$.  This is a result of the
fact that the slope of the stellar mass function below $10^{10.5}$
$\Msun$ in stellar mass (corresponding to $10^{12}$ $\Msun$ in
halo mass) is not steep enough for scatter to have significant impact
\citep[see also][]{Tasitsiomi04}.  Because $\xi>0$ results in
high stellar--mass galaxies being assigned to lower-mass halos than
they would be otherwise (due to the higher number density of
lower-mass halos), the effect is that higher-mass halos contain fewer
stars on average than they would for $\xi=0$.  The effect of setting
$\xi=0$ exceeds systematic error bars only for the very highest mass
halos, above $10^{14.5}$ $\Msun$.

We note that our posterior distribution constrains $\xi$ to be less
than 0.22 dex at the 98\% confidence level.  Higher values for $\xi$
would result in GSMFs inconsistent with the steep falloff of the
\cite{li-2009} GSMF (see also discussion in \citealt{Guo-09}).

\subsubsection{Statistical Errors in Stellar Mass Calculations}

The significance of including or excluding random statistical errors
in stellar mass calculations, $\sigma(z)$,  is also shown Figure
\ref{smmr_sigma_comp}.  The effect of this type of scatter on the
SM--HM relation is mathematically identical to the effect of scatter
in stellar mass at fixed halo mass.  As $\sigma(z=0)$ ($\sim 0.07$
dex) is much smaller than the expected value of $\xi$ ($\sim 0.16$
dex), the convolution of the two effects is only marginally different
from including $\xi$ alone at $z=0$; this results in only a minor
effect on the SM--HM relation.  The effect becomes more pronounced at
$z=1$ for the reason that $\sigma(z=1)$ ($\sim 0.12$ dex) becomes more
comparable to $\xi$---and so including the effects of statistical
errors in stellar mass becomes as important as modeling scatter in
stellar mass at fixed halo mass.

\subsubsection{Cosmology Uncertainties}

\begin{figure}[t!]
\plotgrace{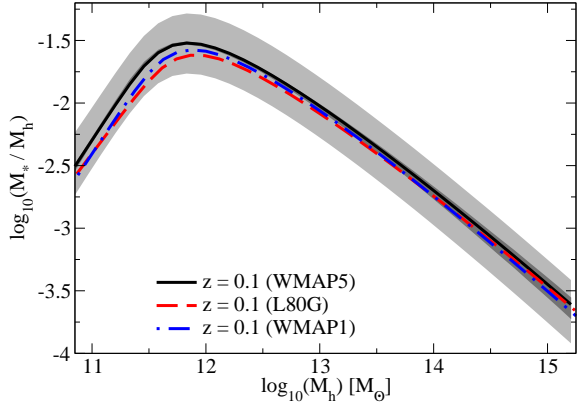}\\
\caption{Comparison between stellar mass fractions in different
  cosmologies.  Light shaded regions denote systematic error spreads
  while dark shaded regions denote error spreads assuming $\mu =
  \kappa = 0$, both about the WMAP5 model.  The dot-dashed blue line
  shows the fiducial relation for a WMAP1 cosmological model (using
  our analytic model) The dashed red line shows the relation for a
  simulation of the WMAP1 cosmology.  Differences between this and the
  analytic model are within the expected sample variance errors.}
\label{smmr_cosmo_comp}
\end{figure}

In Figure \ref{smmr_cosmo_comp}, we show a comparison of best fits for
the stellar mass fraction using abundance matching with three
different halo mass functions: analytic prescriptions for WMAP5 and
WMAP1 (see \S \ref{s:mass_functions}) as well as the mass function
taken directly from the L80G simulation (see \S \ref{simulation}).
The difference between the L80G simulation and the analytic WMAP1 mass
function is slight, as the L80G simulation uses WMAP1 initial
conditions ($h = 0.7$, $\Omega_m = 0.3$, $\Omega_\Lambda = 0.7$, $\sigma_8 = 0.9$, $n_s = 1$);
the difference is consistent with sample variance for
the relatively small (80 $h^{-1}$ Mpc) size of the simulation.  The
difference between SM--HM relations using WMAP1 and WMAP5 cosmologies
is within the systematic errors at all masses.  When systematic errors
are neglected, the two cosmologies yield SM--HM relations that are
noticeably different only at low halo masses ($M<10^{12}M_\odot$).

\begin{figure}[t!]
\plotgrace{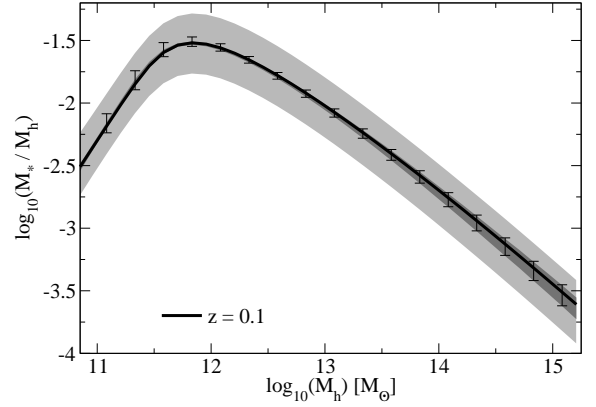}\\
\caption{Effect of cosmological uncertainties on the stellar mass
  fraction at $z=0.1$.  The error bars show the spread in stellar mass
  fractions including both statistical errors and cosmology
  uncertainties (from WMAP5 constraints, \citealt{wmap5}).
  For comparison, the light shaded region includes
  statistical and systematic errors, while the dark shaded region
  includes only statistical errors.}
\label{smmr_cosmo_comp2}
\end{figure}

Figure \ref{smmr_cosmo_comp2} show the results of including
uncertainties in the WMAP5 cosmological parameters.  As described in
\S \ref{s:sys}, this is done using halo mass functions calculated with
parameters resampled from the cosmological parameter chains provided
by the WMAP team.  Only at $z \sim 0$ are the changes in error bars
significant enough to justify mention.  Here, the uncertainty in
cosmology begins to exceed other sources of statistical error for
halos below $10^{12}M_\odot$ due to the small errors on the GSMF at
the stellar masses associated with such halos \citep{li-2009}.
However, the cosmology uncertainties are still well within the
systematic error bars.

\subsubsection{Sample Variance}

\label{s:results_sv}

Because of the large volume of the SDSS, sample variance contributes
insignificantly to the error budget for the SM--HM relation below
$z=0.2$.  Above that redshift, the comparatively limited survey volume
of \cite{perezgonzalez-2007} results in sample variance becoming an
important contributor to the statistical error for halos below
$10^{12} \Msun$ (Poisson noise dominates for larger halos).  If the
effects of sample variance were ignored, the statistical error spreads
for our derived SM--HM relations at $z=1$ would shrink from 0.12 dex
to 0.09 dex for $10^{11} \Msun$ halos, and from 0.05 dex to 0.04 dex
for $10^{12.25} \Msun$ halos.  As with other types of errors, these
considerations are well below the limits of the systematic error bars.

We caution that our error bars including sample variance at $z>0$ have
a very specific meaning.  Namely, they include the standard deviation
in our \textit{fitting form} which might be expected if the survey in
\cite{perezgonzalez-2007} had been conducted on alternate patches of
the sky.  Sample variance at redshifts $z>0$ impacts only the linear
evolution of the SM--HM relations we derive, as the large volume
probed by the SDSS constrains the SM--HM relation very well at $z\sim
0$.  Because our fit is matched to the ensemble of reported data
between $0<z<1$, it is less vulnerable to the effects of sample
variance in individual redshift bins.  Instead, it is affected most by
overall shifts in the number densities reported for the entire
high-redshift survey.  While this means that our fit gives a more
robust SM--HM relation at all redshifts, some caution must be used
when comparing our relation to results derived from the GSMF in a
single redshift bin (e.g., $0.2<z<0.4$).  These will have much larger
uncertainties due to sample variance than SM--HM relations derived
(like ours) from GSMFs along a light cone probing a large redshift range.

To demonstrate the credibility of our approach for calculating the
appropriate error bars including sample variance, we repeated our
analysis of the SM--HM relation using GSMFs from \cite{Drory09} (which
appeared as we were completing this work) instead of
\cite{perezgonzalez-2007} for $0.2 < z < 1$ and retaining the GSMF
from \cite{li-2009} for $z<0.2$.  Although the COSMOS survey in
\cite{Drory09} covers a much larger area ($\sim$9x the area in
\citealt{perezgonzalez-2007}), the fact that it is a single field
means that the expected sample variance is only slightly smaller than
for the combined fields in \cite{perezgonzalez-2007}.  As might be
expected, the SM--HM relation at $z=0.1$ using the \cite{Drory09} GSMF
is identical to the result in Figure \ref{smmr_together_n} because of
the strong constraining power of the SDSS data sample.  At $z=1$, the
SM--HM relation generated by using the \cite{Drory09} GSMF is within
our quoted statistical and sample variance errors, as shown in Figure
\ref{smmr_comparison2}.  This may not be surprising unless one
considers that clustering results suggest an overdensity at the
2-3$\sigma$ level in the COSMOS redshift bin $z=1$ \citep{Meneux09}.
However, because our method fits the \cite{Drory09} GSMFs across the
entire redshift range, the excess at $z=1$ is partially offset by an
underdensity at $z=0.5$.  This demonstrates the robustness of our
fitting method to the effects of sample variance except on the scale
of the entire survey.

\subsubsection{Satellite Treatment}

\label{s:sat_results}

\begin{figure}[t!]
\plotgrace{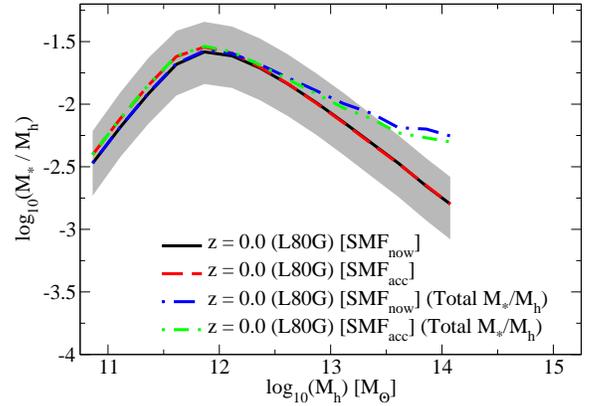}
\caption{Comparison between stellar mass fractions and total stellar
  mass fractions (labeled as ``Total M$_\ast$/M$_\mathrm{h}$'')
  derived by assuming different matching epochs for satellite
  galaxies.  The L80G simulation was used here in order to follow the
  accretion histories of the subhalos.  The relations terminate at
  high masses where the halo statistics become unreliable due to
  finite--volume effects.}
\label{smmr_z_acc_comp}
\end{figure}

Finally, we consider the changes in both the stellar mass fraction and
total stellar mass fraction (total stellar mass in central galaxy and
all satellites / total halo mass) induced by different satellite
evolution models (see \S \ref{s:sats} for details on the two models).
As shown in Figure \ref{smmr_z_acc_comp}, fixing satellite stellar
mass at the redshift of accretion (lines labeled as
``SMF$_\mathrm{acc}$'') has virtually no effect on either fraction as
compared to allowing satellite stellar mass to evolve the same way as
centrals with the same mass (labeled as SMF$_\mathrm{now}$).  Because
comparison of different satellite evolution models requires tracking
satellites through merger trees, Figure \ref{smmr_z_acc_comp} shows
results only for satellites in the L80G simulation.

The treatment of satellites may have a somewhat larger impact on the
total stellar mass fraction, including the stellar mass of both
central and satellite galaxies within a halo.  This is shown for both
models in Figure \ref{smmr_z_acc_comp}.  Because of the steep fall-off
in stellar mass for low mass galaxies, the total stellar mass fraction
has only minimal contribution from satellites for low mass halos, and
deviates significantly from the stellar mass fraction for centrals
only at halo masses $M_h > 10^{12.5} \Msun$.  At cluster-scale masses
($M_h \sim 10^{14} \Msun$), accreted satellites have on average a
higher ratio of stars to dark matter than the central galaxy, and the
total stellar mass fraction can be many times the central stellar mass
fraction.  However, the impact of the two models for satellite
treatment on this ratio is small.  Profiles of satellite galaxies in
clusters should be able to better distinguish between such models.

\subsubsection{Summary of Most Important Uncertainties}

\label{s:uncertainties_summary}

Systematic stellar mass offsets resulting from modeling choices result
in the single largest source of uncertainties ($\sim$0.25 dex at all
redshifts).  The contribution from all other sources of error is much
smaller, ranging from 0.02-0.12 dex at $z=0$ and from 0.07-0.16 dex at
$z=1$.  On the other hand, this statement is only true when all
contributing sources of scatter in stellar masses are considered.
Models that do not account for scatter in stellar mass at fixed halo
mass will overpredict stellar masses in $10^{14.25}$ $\Msun$ halos by
0.13-0.19 dex, depending on the redshift.  Models that do not account
for scatter in calculated stellar mass at fixed true stellar mass will
overpredict stellar masses in $10^{14.25}$ $\Msun$ halos by 0.12 dex
at $z=1$.  Hence, it is important to take both these effects into
account when considering the SM--HM connection either at high masses
or at high redshifts.

\subsection{Comparison with other work}
\label{s:comparison}

\begin{figure*}
\plotlargegrace{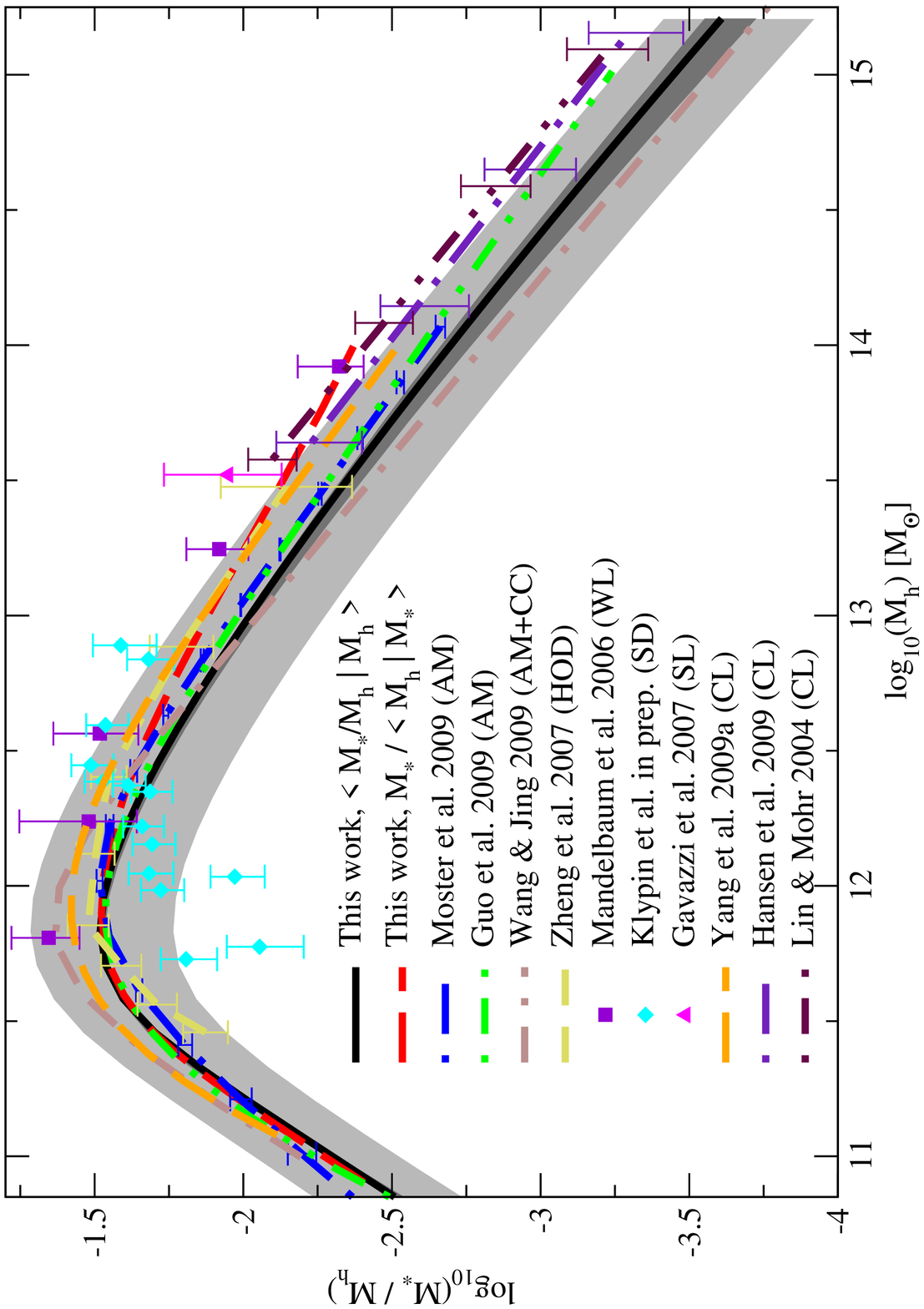}
\caption{Comparison of our best-fit model at $z=0.1$ to previously
  published results.  Results shown include other results from
  abundance matching (\citealt{moster-09} and \citealt{Guo-09}); abundance
  matching plus clustering constraints \citep{Wang09}; HOD
  modeling \citep{zheng-07}; direct measurements from weak lensing
  \citep{mandelbaum-06}, statellite dynamics \citep{klypin09} and strong
  lensing \citep{Gavazzi07}; and clusters selected from SDSS spectroscopic
 data \citep{yang-08}, SDSS photometric data (the maxBCG sample
  \citealt{Hansen09}), and X-ray selected clusters \citep{LinMohr04}.
  Dark grey shading indicates statistical and sample variance errors;
  light grey shading includes systematic errors.  The red line shows
  our results averaged over stellar mass instead of halo mass; scatter
  affects these relations differently at high masses.  The results of
  \citet{mandelbaum-06} and \citet{klypin09} are determined by
  stacking galaxies in bins of stellar mass, and so are more appropriately
  compared to this red line.}
\label{smmr_comparison}
\end{figure*}

A comparison of our results with several results in the literature at
$z\sim 0.1$ is shown in Figure \ref{smmr_comparison}.  Such comparison
is not always straightforward, as other papers have often made
different assumptions for the cosmological model, the definition of
halo mass, or the measurement of stellar mass.  In addition, some
papers report the average stellar mass at a given halo mass (as we
do), and others report the average halo mass at a given stellar mass.
Given the scatter in stellar mass at fixed halo mass, the averaging
method can affect the resulting stellar mass fractions, particularly
for group- and cluster-scale halo masses.  To facilitate comparison
with both approaches, we plot our main results (labeled as ``$\langle
M_\ast/M_h | M_h\rangle$'') along with results for which the stellar
mass fractions have been averaged at a given stellar mass (labeled as
``$\langle M_\ast/M_h | M_\ast\rangle$'').  In the comparisons below,
we have not adjusted the assumptions used to derive stellar masses,
because such adjustments can be complex and difficult to apply using
simple conversions.  Additionally, we have only corrected for differences
in the underlying cosmology for those papers using a variant of
abundance matching method \citep{moster-09,Guo-09,Wang09,Conroy09}
using the process described in Appendix
\ref{direct_matching}, as alternate methods require corrections which
are much more complicated.  We have, however, adjusted the IMF of
all quoted stellar masses to that of \citet{chabrier-2003-115}, and we have
converted all quoted halo masses to virial masses as defined in \S \ref{s:mass_functions}.

The closest comparison with our work, using a very similar method, is
the result from \cite{moster-09}.  This result is in excellent
agreement with ours at the high mass end, and is within our systematic
errors for all masses considered.  However, their less flexible choice
of functional form, and their use of a different stellar mass function
(estimated from spectroscopy using the results of \citealt{panter:2007}) results in a
different value for the halo mass $M_{peak}$ with peak stellar mass
fraction and a shallower scaling of stellar mass with halo mass at the
low mass end.  Their error estimates only account for statistical
variations in galaxy number counts, and they do not include sample
variance or variations in modeling assumptions.  \cite{Guo-09} use a
similar approach to \cite{moster-09}, using stellar masses from
\cite{li-2009}, but they do not account for scatter in stellar mass at
fixed halo mass.  Consequently, their results match ours for $10^{12}$
$\Msun$ and less massive halos, but overpredict the stellar mass for
larger halos.

\cite{Wang09} use a parameterization for the SM--HM relation for both
satellites and centrals, and they attempt to simultaneously fit both
the stellar mass function and clustering constraints, including the
effects of scatter in stellar mass at fixed halo mass.  At $z\sim
0.1$, their data source matches ours \citep{li-2009}, but their
approach finds a best-fit scatter
  in stellar mass at fixed halo mass of $\xi = 0.2$ dex, essentially
  the highest value allowed by the stellar mass function
  \citep{Guo-09}. As this is higher than our best-fit value for $\xi$,
  their SM--HM relation falls below ours for high-mass galaxies.
  Possibly because of the limited flexibility of their fitting form
  (they use only a four-parameter double power-law), their SM--HM
  relation is in excess of ours for halo masses near $10^{12} \Msun$.

  \cite{zheng-07} used the galaxy clustering for luminosity-selected
  samples in the SDSS to constrain the halo occupation distribution.
  This gives a direct constraint on the $r-$band luminosity of central
  galaxies as a function of halo mass.  Stellar masses for this sample
  were determined using the $g-r$ color and the $r$-band luminosity as
  given by the \cite{bell:2003} relation, and a WMAP1 cosmology was
  assumed.  This method allows for scatter in the luminosity at fixed
  halo mass to be constrained as a parameter in the model; results for
  this scatter are consistent with \cite{more-09}, although they are
  less well constrained.  According to \cite{li-2009}, stellar masses
  for the \cite{bell:2003} relation are systematically larger than
  those calculated using \cite{blanton-roweis-07} by 0.1--0.3 dex.
  However, as $\Omega_m$ in WMAP1 is larger than in WMAP5, halo masses
  in WMAP1 will be higher at a given number density than in WMAP5,
  somewhat compensating for the higher stellar masses.

We next compare to constraints from direct measurements of halo masses
from dynamics or gravitational lensing.  \cite{mandelbaum-06} have
used weak lensing to measure the galaxy--mass correlation function for
SDSS galaxies and derive a mean halo mass as a function of stellar
mass.  \cite{mandelbaum-06} assume a WMAP1 cosmology and uses
spectroscopic stellar masses, calculated per \cite{Kauffmann03a}.
Klypin et al (in preparation) have derived the mean halo mass as a function of
stellar mass using satellite dynamics of SDSS galaxies \citep[see
also][]{Prada03, VDB04b, conroy-07}.  Their results are generally
within our systematic errors but lower than others at the lowest
masses and with a somewhat different shape.  This may be due to
selection effects, as their work uses only isolated galaxies, which may
have somewhat lower average stellar masses. 
 \cite{Gavazzi07} use a set of strong lenses from the SLACS survey
along with a model for simultaneously fitting the stellar and
dark matter components of the stacked lens profiles.  This result, at
one mass scale, is a bit higher than our error range but within 1.5
$\sigma$.  The selection effects relevant to strong lenses are beyond
the scope of this paper; however, within the effective radius, the
stellar mass can easily contribute more to the lensing effect than the
dark matter.  Thus, at any given halo mass, the halos with less
massive galaxies are much less likely to be strong lenses, resulting
in a bias towards higher stellar mass fractions in strong lenses as
compared to halos selected at random.

At the high mass end, one can directly identify clusters and groups
corresponding to dark matter halos, and measure the stellar masses of
their central galaxies.  \cite{yang-08} use a group catalog matched to
halos to determine halo masses (via an iteratively-computed group
luminosity--mass relation).  Stellar masses in this work are
determined using the \cite{bell:2003} relation between $g-r$ color and
$M/L$; a WMAP3 cosmology was assumed.  Their results agree very well
with ours for low-mass halos, but they begin to differ at higher
masses.  This may be partially due to scatter between their calculated
halo masses (based on total stellar mass in the groups) and the true
halo masses, resulting in additional scatter in their stellar masses
at fixed halo mass.  It could also be due to differences in stellar
modeling; their results remain at all times within our systematic
errors.  We also compare to direct measurements of massive clusters by
\cite{Hansen09} and \cite{LinMohr04}.  In order to convert
luminosities to stellar masses, we assume $M/L_{i0.25}$ =
3.3$M_\Sun/L_{\Sun,i0.25}$ and $M/L_{K}$ = 0.83$M_\Sun/L_{\Sun,K}$
based on the population synthesis code of \citet{Conroy09}.  These
measurements are both somewhat higher than our results for massive
clusters, the one-sigma error estimates overlap.  The discrepancies
may be due to issues with cluster selection and with modeling scatter
in the mass-observable relation; in each case the cluster mass is an
average mass for the given observable (X-ray luminosity or cluster
richness), and can result in a bias if central galaxies are correlated
with this observable.  More detailed modeling of the scatter and
correlations will be required to determine whether this is can account for
the offsets.

\begin{figure}
\plotgrace{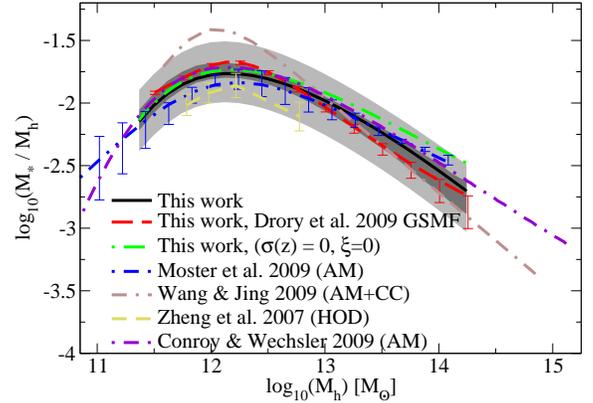}
\caption{Comparison of our best-fit model at $z=1.0$ for different model assumptions
  and to previously published results.  Dark grey
  shading indicates statistical and sample variance errors; light grey
  shading includes systematic errors.  The error bars for the red
  line, calculated using the \cite{Drory09} GSMF include statistical
  errors only---i.e., they do not include sample variance.  The
  results of \citet{moster-09} (green line) do not include modeling of
  scatter or statistical errors in stellar masses, so for comparison,
  we present our results excluding the effects of $\sigma(z)$ and
  $\xi$ (blue line).  The results of \cite{cw-08} made
  slightly different assumptions about the stellar mass function
  evolution.}
\label{smmr_comparison2}
\end{figure}

A comparison of our results to others at $z\sim1$ is shown in Figure
\ref{smmr_comparison2}.  As may be expected, it is much harder to
directly measure the SM--HM relation at higher redshifts, resulting in
relatively fewer published results with which we may compare.  We
first note that we have compared the impact of two independent
measurements of the GSMF from different surveys.  As discussed in
\ref{s:results_sv}, because we simultaneously fit our model with
linear evolution to the GSMF at redshifts $0 < z < 1$, our results are
less sensitive to sample variance.  In contrast to the conclusion of
\cite{Drory09} which fit their results to specific redshift bins,
Figure \ref{smmr_comparison2} shows that the \cite{Drory09} and
\cite{perezgonzalez-2007} results are in agreement within statistical
errors at $z \sim 1$ when fitting the full redshift range. The results
in \cite{moster-09} are also very similar to ours at $z\sim 1$.
However, \cite{moster-09} do not give fits for the SM--HM relation
which include the effects of scatter in stellar mass at fixed halo
mass except at $z\sim 0$ and do not in general include scatter in
measured stellar mass with respect to the true stellar mass.  Hence,
we include for comparison an SM--HM relation derived using our
analysis but excluding both of these effects.  The remaining deviance
most likely stems from sample variance due to the much smaller
survey volume on which their fit is based.

At $z \sim 1$, \cite{Wang09} make use of clustering data and stellar
mass functions from the VVDS survey \citep{Meneux08,Pozzetti07} at
$z\sim 0.8.$ At the same time, the high-mass and low-mass slopes of
their power-law relation are not re-fit for the higher redshift,
leading to similar deviations from our results as at $z\sim0.1$.  The
results in \cite{zheng-07} lie slightly below our results at $z\sim
1$.  This is partially due to their use of a WMAP1 cosmology.
However, it is difficult to say exactly how much, as their stellar
masses at $z\sim 1$ are a hybrid of $K_s$-derived masses from
\cite{Bundy06} and color--based masses derived in a manner analogous
to \cite{bell:2003}.  Nonetheless, they remain well within our
systematic error bars. 

We also include a comparison to the $z=1$ SM--HM relation
presented in \cite{cw-08}, who also use an abundance matching
technique to assign galaxies to halos.  Unique to the work of
\cite{cw-08} is their attempt to jointly fit both the
redshift--dependent stellar mass functions and the redshift--dependent
star formation rate -- stellar mass relations.  In their model, halo
growth is tracked through time using results derived from halo merger
trees, allowing galaxies to be identified across epochs.  The
evolution of halos in conjunction with standard abundance matching
provides model predictions for star formation rates.  The SM--HM
relation from \cite{cw-08} lies slightly above our best--fit relation, and
it is within statistical error bars except at the very highest halo masses.
This is due to the different assumptions made about the GSMF in the
two works and also to the absence of corrections for the scatter in
stellar mass at fixed halo mass in \cite{cw-08}.

Finally, the strong lensing survey of \cite{Gavazzi07} has been
extended by \citep{Lagattuta09} out to $z \sim 0.9$ using lenses
observed in the CASTLES program, as well as in COSMOS and in the EGS.
While the same caveats about selection effects apply as for lower
redshifts, \cite{Lagattuta09} find that the evolution in the stellar
mass fraction for $M_h \sim 10^{13.5} \Msun$ halos is within 0-0.3 dex
greater at $z\sim 0.9$ as compared to $z \sim 0$, consistent with our
limits of 0-0.15 dex for the allowed evolution over that redshift
interval.

\section{Results Beyond $\lowercase{z}=1$}
\label{s:results_highz}

\subsection{Methodology and Data Limitations}

As discussed in \S \ref{SFRs}, published results for the galaxy
stellar mass function beyond $z=1$ suffer from the important caveat
that integrated SFRs are inconsistent with galaxy stellar mass
functions when both sets of observations are taken at face value with
a constant IMF.  Nonetheless, one may use similar methodology as in \S
\ref{s:methodology} to derive stellar mass -- halo mass relations at
higher redshifts under the assumption that the observed stellar mass
functions are correct. Here we assume the GSMFs of
\cite{marchesini-2008}, which cover a redshift range of $1.3 < z < 4$.

As there is no guarantee that the evolution of the SM -- HM relation
at high redshifts will have the same form as its evolution at low
redshifts, we re-examine the assumptions affecting our evolution
parameterization in Equation \ref{eq:zscaling}.  As with all current
high-redshift data, the results in \cite{marchesini-2008} are limited
to luminous (massive) galaxies, so little information about the value
of $\beta$ (the faint-end slope of the galaxy-halo mass relation) is
available.  Hence, we continue to assume a linear functional form for
its evolution; as the value of $\beta$ evolves linearly with scale
factor ($a$) in our fit, this means that it is largely constrained to
be consistent with the evolution at lower redshifts ($1<z<2$).
Naturally, if the evolution of $\beta$ at high redshifts is
significantly different than for $1<z<2$, then our error bars for
$\beta$ may underestimate the full uncertainties in the parameter.

Additionally, the systematics affecting high-mass galaxies at high
redshifts are much more severe than for $z<1$.  Not only are the
errors in stellar mass calculations significant (due to larger
photometry errors, limited templates, etc.), but any miscalibration in
correcting photometric redshift errors will result in low-redshift
galaxies masquerading as very bright high-redshift galaxies.  These
combined uncertainties result in poor constraints on high-mass
galaxies.  For that reason, we do not attempt to assume a more
complicated functional form for the evolution of $\delta$ and
$\gamma$, which means as before that their rates of evolution are
largely constrained to be consistent with lower redshifts.

However, we find that individual fits at each redshift do suggest a
possible evolution for the characteristic stellar mass which is
nonlinear in the scale factor.  Hence, we expand the form of the
evolution of $M_{\ast,0}$ to include a quadratic dependence on scale
factor:
\begin{equation}
M_{\ast,0}(a) = M_{\ast,0,0} + M_{\ast,0,a}(a - 1) + 
M_{\ast,0,a^2}(a-0.5)^2,
\label{eq:fit2}
\end{equation}
where $(a - 0.5)^2$ is used instead of $(a-1)^2$ to minimize the
degeneracy between $M_{\ast,0,a}$ and $M_{\ast,0,a^2}$.  We
parameterize the evolution of all other parameters as in Equation
\ref{eq:zscaling}.  All other methodology remains the same as for
lower redshifts, as outlined in \S \ref{s:summary}.

\subsection{Results}
\label{s:discussion_highz}

To maintain some overlap with the $z<1$ results, we evaluate the
likelihood function for each SM -- HM relation against GSMFs for $0.8
< z < 4$.  This data range includes two redshift bins from
\cite{perezgonzalez-2007} (0.8-1.0, 1.0-1.3) and three redshift bins
from \cite{marchesini-2008} (1.3-2, 2-3, 3-4).  Including more
redshift bins from \cite{perezgonzalez-2007} would improve the
continuity of the fits to the low-redshift results; however, doing so
would require a more complicated redshift parameterization than what we
have assumed.  The evolution of the best-fit stellar mass fraction for
$0<z<4$ is shown in Figure \ref{all_cen_together_z4_00}.  All data
points for Figure \ref{all_cen_together_z4_00} are listed in Appendix
\ref{a_data}.

As may be expected, uncertainties at high redshifts are substantially
larger than at lower redshifts.  The contribution of systematic errors
in stellar masses to the error budget (0.25 dex) is still important,
but it is no longer the only dominant factor.  Statistical errors due
to the comparatively small number of galaxy observations at high
redshifts can contribute an equal uncertainty (up to 0.25 dex) to the
derived SM--HM relation.  The statistical errors are large not only
for massive galaxies with low number counts, but also for halos below
$10^{12}$ $\Msun$, where magnitude limits on surveys make observations
of the corresponding galaxies difficult.

The contribution from other sources of uncertainties (e.g., sample
variance, cosmology uncertainties) is substantially smaller than the
current statistical errors.  The effects of sample variance on
uncertainties at high redshift is less than for low redshifts because
the volume probed in the high redshift sample is five times larger
than for the low redshift sample ($\approx 5\times10^6$ Mpc$^3$ vs.
$10^6$ Mpc$^3$, respectively).  While cosmology uncertainties are
somewhat larger at high redshifts, their contribution to the overall
levels of uncertainty are again much smaller than the statistical
errors, as was true even by $z=1$ for the low-redshift sample.

The statistical uncertainties at high redshifts mean that it is
difficult to draw strong conclusions about the evolution of the SM--HM
relation.  However, the indication is that the mass corresponding to
the peak efficiency for star formation evolves slowly, and is roughly
a factor of five larger at $z=4$.  At all redshifts, the integrated
star formation peaks at $\sim$ 10-20 per cent of the universal baryon
fraction; the current data indicates that this value may start high at
very high redshifts, shrink as halos grow faster than they form stars,
and then start growing again after $z=2$.  However, with current
uncertainties these results are tenuous.  The single most effective
way to reduce current uncertainties on both the SM--HM relation at
individual redshifts and on its evolution is to conduct more
high-redshift galaxy surveys, both to probe fainter galaxies to
determin the shape of the GSMF and to get better statistics at the
high mass end.

\begin{figure}[!t]
\plotgrace{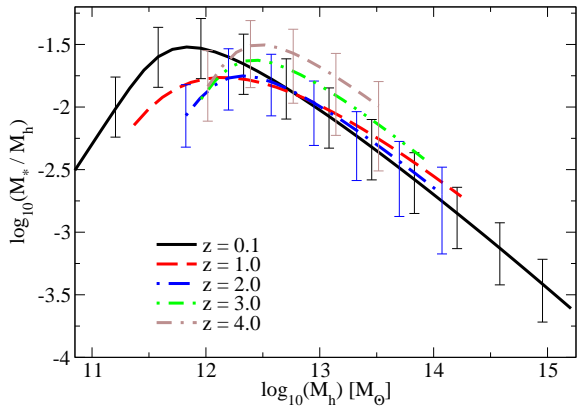}
\caption{Evolution of the derived stellar mass fractions for central
  galaxies, from $z=4$ to the present.  The best--fit relations are
  shown only over the mass range where constraining data are
  available.  At higher redshifts, certainty about the shape of the
  curves drops precipitously owing to a lack of constraining data
  beyond the knee of the stellar mass function.  Combined systematic
  and statistical error bars are shown for three redshift bins only.}
\label{all_cen_together_z4_00}
\end{figure}

\section{Discussion and Implications}
\label{s:discussion}

At $z\sim 0$, the majority of published results are in accord within
our full systematic error bars, regardless of the technique used.  All
reported results appear to be consistent with the principles necessary for
abundance matching over a wide range of halo masses ($10^{11}-10^{15}$
$\Msun$)---that each dark matter halo and subhalo above the masses we
have considered hosts a galaxy with a reasonably tight relationship
between their masses, and that average stellar mass --- halo mass relation
increases monotonically with halo mass.

Because of the available statistics of halo and galaxy stellar mass
functions, especially at $z=0$, the technique of abundance matching
offers the tightest constraints on the SM--HM relation currently
available, and it is in agreement with results from a broad variety of
additional techniques.  Under the assumption that systematic errors in
stellar mass calculations do not change substantially with redshift,
abundance matching offers tight constraints on the evolution of the
SM--HM relation from $z=1$ to the present.  These in turn will serve
as important new tests for star formation prescriptions and recipes in
both hydrodynamical simulations and semi-analytic models, as they will
apply on the level of individual halos instead of on the simulated
volume as a whole.

At the same time, abundance matching offers these constraints with a
minimal number of parameters.  The Halo Occupation Distribution (HOD)
technique requires modeling $P(N|M_h)$, the probability distribution
of the number of galaxies per halo as a function of halo mass, in
several different luminosity bins.  In the model proposed in
\citet{zheng-07}, this results in 45 fitted parameters just to models
the occupation at $z=0$ (five parameters for nine luminosity bins).
Conditional Luminosity Function (CLF) modeling requires parameterizing
a form for $\phi(L,M_h)$, the number density of galaxies as a function
of luminosity and host halo mass, which results in approximately a
dozen parameters to model occupation at $z=0$ \citep{Cooray06}.
Because of the additional constraints imposed by assuming that each
halo hosts a galaxy, our approach uses fewer parameters.  Abundance
matching, as discussed in this paper, results in a model with only six
independent parameters (five to empirically fit the derived SM--HM
relation, and one to model the scatter in observed stellar masses at
fixed halo mass) to describe the population of galaxies in halos.

The abundance matching approach to the SM--HM relation requires so few
parameters in comparison to other methods because of the fairly small
scatter ($\approx 0.16$ dex) between stellar mass and halo mass at
high masses (the scatter has a negligible impact on the average SM--HM
relation at lower masses), and the requirement that satellite galaxies
live in satellite halos (subhalos).  It may well be that a more
complicated model must be adopted for satellites to quantitatively
match the small-scale clustering observations \citep[e.g.][]{Wang06}.
However, such changes will affect the clustering much more than the
derived SM--HM relation, as suggested by the minimal changes in Figures
\ref{smmr_sigma_comp} and \ref{smmr_z_acc_comp} for mass scales
($\lesssim 10^{12.5}$ $\Msun$) where satellites are a non-negligible
fraction of the total halo population.

The largest uncertainties in the SM --- HM relation at $z<1$ come from
assumptions in converting galaxy luminosities into stellar masses,
which amount to uncertainties on the order of $0.25$ dex in the
normalization of the relation.  However, the systematic biases
introduced by the combined sources of scatter between calculated
stellar masses and halo masses can rise to equivalent significance for
halos above $10^{14.5}$ $\Msun$.  Because the GSMF is monotonically
decreasing, results which do not account for all sources of scatter in
stellar mass will overpredict the average stellar mass in halos by
0.17-0.25 dex for these massive halos.

Using abundance matching to find confidence intervals for the SM--HM relation is an even more involved process, as each of the ways in
which the systematics might vary must also be taken into
account. While future work on constraining stellar masses will be the
most valuable in terms of reducing uncertainties for the lowest redshift
data, wider and deeper surveys and some resolution to the discrepancy
between high-redshift cosmic star formation density and stellar mass
functions must occur in order to improve constraints on the relation
at high redshifts.

As mentioned in the introduction, abundance matching may be used
equally well to assign galaxy luminosities and colors to halos.  In
this case, the galaxy luminosity --- halo mass relation may be derived
using identical methodology to that presented in \S
\ref{s:methodology}, with the exception that the systematics $\mu$ and
$\kappa$ may be neglected, leaving only $\sigma(z)$ (effectively, the
redshift scaling of photometry errors) and $\xi$ (effectively, the
scatter in luminosity at fixed halo mass).  As these systematics are
much better constrained than their stellar mass counterparts (simply
as luminosities may be measured directly), this approach can yield
very powerful constraints on the normalization as well as the
evolution of the luminosity--mass relation.  The higher accuracy
possible compared to the stellar mass--mass relation will generally
\textit{not} remove uncertainties in comparing to galaxy formation models.
Galaxy formation codes which calculate luminosities must include modeling
for all the effects in \S \ref{syst_stellarm}, meaning that constraints on the
underlying physics are subject to the same uncertainties.
However, tighter constraints on the luminosity--mass relation will be
nonetheless helpful for applications which are concerned with cosmological
constraints from large luminosity-selected surveys.

\section{Conclusions}
\label{s:conclusions}

We have performed an extensive exploration of the uncertainties
relevant to determining the relationship between dark matter halos and
galaxy stellar masses from the halo abundance matching technique.
Errors related to the observed stellar mass function, the theoretical
halo mass function, and the underlying technique of abundance matching
are all considered.  We focus on the mean stellar mass to halo mass
ratio for central galaxies as a function of halo mass, and present
results for this relationship at the present epoch and extending
to $z \sim 4$.  We account separately for statistical errors and for
systematic errors resulting from uncertainties in stellar mass
estimation, and also investigate the relative contribution of various
sources of error including cosmological uncertainty and the scatter
between stellar mass and halo mass.  An analytic model has been
developed which can be used to constrain this connection in the
absence of high resolution simulations.

Our primary conclusions are as follows:

\begin{enumerate}
\item The peak integrated star formation efficiency occurs at a halo
  mass near $10^{12}$ $\Msun$, with a relatively low fraction --- 20\%
  at $z=0$ --- of baryons currently locked up in stars.  This peak
  value declines to $z\sim 2$ but remains between 10--20\% for all
  redshifts between $z = $0--4.  This implies that $30-40$\% of
  baryons were converted into stars over the lifetime of a galaxy with
  current halo mass of $10^{12}$ $\Msun$.
\item At low masses, the stellar mass -- halo mass relation at $z=0$ scales as
  $M_\ast \sim M_h^{2.3}$.  At high masses, around $10^{14}$
  $\Msun$, stellar mass scales as $M_\ast \sim M_h ^{0.29}$.
  However, the high mass scaling may not be a power law, as our model
  indicates that this slope decreases with increasing halo mass.
\item Within statistical uncertainties, the stellar mass content of
  halos has increased by $0.3-0.45$ dex for halos with mass less than
  $10^{12}$ $\Msun$ since $z \sim 1$.  Systematic biases in stellar
  mass calculations between different redshifts could broaden the
  uncertainties in this number, but the conclusion that significant
  evolution has occurred for low-mass halos would remain robust.  For
  halos with mass greater than $10^{13} \Msun$, our best-fit results
  indicate more growth in halo masses than stellar masses since $z
  \sim 1$, but are consistent with no evolution in the stellar mass
  fractions over this time.
\item Systematic, uniform offsets in the galaxy stellar mass function
  and its evolution are the dominant uncertainty in the stellar mass
  -- halo mass relation at low redshift.  Statistical errors in the
  estimation of individual stellar masses impact the high mass end of
  the GSMF, and at higher redshifts may result in an observed GSMF
  that deviates from a Schechter function.
\item Current uncertainties in the underlying cosmological model are
  sub-dominant to the systematic errors, but are larger than other
  sources of statistical error for halos below $M_h \sim 10^{12}$
  $\Msun$ for low redshifts ($z<0.2$).
\item Given current constraints from other methods, uncertainty in the
  value of scatter between stellar mass and halo mass is important in
  the mean relation for masses above $M_h \sim 10^{12.5}$ $\Msun$,
  although it is subdominant to systematic errors for all masses below
  $M_h \sim 10^{14.5}$ $\Msun$.
\item Other uncertainties in the galaxy--halo assignment, including
  different assumptions about the treatment of satellite galaxies, are
  subdominant when considering the mean relation for central galaxies.
\item At higher redshifts ($1 < z < 4$), systematic uncertainties remain
  important, but statistical uncertainties reach equal significance.
  The shape of the relation is fairly unconstrained at $z>2$, where
  improved statistics and constraints on the GSMF below $M*$ are
  needed.
\end{enumerate}

We have presented a best--fit galaxy stellar mass -- halo mass
relation including an estimate of the total statistical and systematic
errors using available data from $z=0-4$, although caution should be
used at redshifts higher than $z \sim 1$.  We also present an
algorithm to generalize this relation for an arbitrary cosmological
model or halo mass function.  The fact that assignment errors are
sub-dominant and scatter can be well--constrained by other means gives
increased confidence in using the simple abundance matching approach
to constrain this relation.  These results provide a powerful
constraint on models of galaxy formation and evolution.

\acknowledgments

PSB and RHW received support from the U.S. Department of Energy under
contract number DE-AC02-76SF00515 and from a Terman Fellowship at
Stanford University.  CC is supported by the Porter Ogden Jacobus
Fellowship at Princeton University.  We thank Michael Blanton, Niv
Drory, Raphael Gavazzi, Qi Guo, Sarah Hansen, Anatoly Klypin,
Cheng Li, Yen-Ting Lin, Pablo P\'erez-Gonz\'alez, Danilo Marchesini,
Benjamin Moster, Lan Wang, Xiaohu Yang, Zheng Zheng, as well as their co-authors
for the use of electronic versions of their data.  We appreciate many helpful
discussions and comments from Ivan Baldry, Michael Busha, Simon Driver,
Niv Drory, Anatoly Klypin, Ari Maller, Danilo Marchesini, Phil Marshall, Pablo
P\'erez-Gonz\'alez, Paolo Salucci, Jeremy Tinker, Frank van den Bosch, 
the Santa Cruz Galaxy Workshop, and the anonymous referee for this paper.
The ART simulation (L80G)
used here was run by Anatoly Klypin, and we thank him for allowing us
access to these data.  We are grateful to Michael Busha for providing the
halo catalogs we used to estimate sample variance errors. These
simulations were produced by the LasDamas project
( {\tt http://lss.phy.vanderbilt.edu/lasdamas/} ); we thank the LasDamas
collaboration for providing us with this data.

\bibliography{master_bib}

\begin{appendix}

\section{Converting Results to Other Halo Mass Functions}
\label{direct_matching}

From Equation \ref{direct_matching_eqn}, it is possible to simply convert
from our halo mass function $\phi_\mathrm{h}$ to any halo mass
function of choice ($\phi_\mathrm{h,r}$).  In particular, the
function $M_\mathrm{h}(M_\ast)$ is defined by the fact that the number
density of halos with mass above $M_\mathrm{h}(M_\ast)$ must match the
number density of galaxies with stellar mass above $M_\ast$ (with the
appropriate deconvolution steps applied).  Recall from Equation
\ref{direct_matching_eqn} that

\begin{equation}
\int_{M_\mathrm{h}(M_\ast)}^\infty \phi_\mathrm{h}(M)d\log_{10} M = \int_{M_\ast}^\infty \phi_\mathrm{direct}(M_\ast)d\log_{10} M_\ast.
\end{equation}

Naturally, the correct mass-stellar mass relation for the alternate
halo mass function $\phi_\mathrm{h,r}$
(which we will label as $M_\mathrm{h,r}(M_\ast)$) must
satisfy this same equation, with the result that:
\begin{equation}
\int_{M_h(M_\ast)}^\infty \phi_\mathrm{h}(M)d\log_{10} M = \int_{M_\mathrm{h,r}(M_\ast)}^\infty  \phi_\mathrm{h,r}(M)d\log_{10} M.
\end{equation}

To make the calculation even more explicit, let $\Phi_\mathrm{h}(M) = \int_M^\infty \phi_\mathrm{h} d\log_{10}M$ be our cumulative halo mass function, and let $\Phi_\mathrm{h,r}(M)$ be the corresponding cumulative halo mass function for $\phi_{h,r}$.  Then, we find:
\begin{equation}
M_\mathrm{h,r}(M_\ast) = \Phi_\mathrm{h,r}^{-1}(\Phi_\mathrm{h}(M_\mathrm{h}(M_\ast))).
\end{equation}

Mass functions from different cosmologies
than those assumed in this paper will also require converting stellar masses
if their choices of $h$ differ from the WMAP5 best-fit value.

\section{Effects of Scatter on the Stellar Mass Function}

\label{scatter_effects}

This section is intended to provide basic intuition for the effects of
both $\xi$ and $\sigma(z)$, which may be modeled as convolutions.
The classic example in this case is convolution of the GSMF with a
log-normal distribution of some width $\omega$.  While the convolution
(even of a Schechter function) with a Gaussian has no known analytical
solution, we may approximate the result by considering the case where
the logarithmic slope of the GSMF changes very little over the width of
the Gaussian.  Then, locally, the stellar mass function is
proportional to a power function, say $\phi(M_\ast) \propto
(M_\ast)^\alpha$.  Then, if we let $x = \log_{10} M_\ast$ (so that
$\phi(10^x) \propto 10^{\alpha x}$), finding the convolution is
equivalent to calculating the following integral:
\begin{eqnarray}
\phi_\mathrm{conv}(10^x) &\propto& \int_{-\infty}^\infty \frac{10^{\alpha b}}{\sqrt{2\pi\omega^2}} \exp\left(-\frac{(x-b)^2}{2\omega^2}\right)db \nonumber\\
& = & 10^{\alpha x} \, 10^{\frac{1}{2}\alpha^2 \omega^2 \ln(10)}.
\end{eqnarray}
That is to say, the stellar mass function is shifted upward by
approximately $1.15\,(\alpha\omega)^2$ dex.  Hence, for parts of
the stellar mass function with shallow slopes, the shift is completely
insignificant, as it is proportional to $\alpha^2$.  However, it
matters much more in the steeper part of the stellar mass function, to
the point that for galaxies of mass $10^{12}$ $\Msun$, the observed stellar mass
function can be several orders of magnitude above the intrinsic GSMF.

\section{A Sample Calculation of the Functional Form of the Stellar
  Mass Function}

\label{sample_smf_calc}
Galaxy formation models typically assume at
least two feedback mechanisms to limit star formation for low-mass
galaxies and for high-mass galaxies. Thus, one of the simplest
fiducial star formation rate (SFR) as a function of halo mass
($M_h$) would assume a double power-law form:

\begin{equation}
SFR(M_h) \propto \left(\frac{M_h}{M_0}\right)^a + \left(\frac{M_h}{M_0}\right)^b.
\end{equation}

We might expect the total stellar mass as a function of halo mass to
take a similar form, except perhaps with a wider region of transition
between galaxies whose histories are predominantly low mass, and those
with histories which are predominantly high mass, for the reason that
some galaxies' accretion histories may have caused them to be affected
comparably by both feedback mechanisms.

Hence, assuming that the relation between halo mass and stellar mass
follows a double power-law form, we adopt a simple functional form to
convert from the stellar mass of a galaxy to the halo mass:
\begin{equation}
\label{double_power_law_relation}
M_h(M_\ast) = M_1 \, \left[\left(\frac{M_\ast}{M_{\ast,0}}\right)^{\beta/\gamma} + \left(\frac{M_\ast}{M_{\ast,0}}\right)^{\delta/\gamma}\right]^\gamma.
\end{equation}
Here, $\beta$ may be thought of as the faint-end slope, $\delta$ as
the massive-end slope (although $\beta$ and $\delta$ are functionally
interchangeable), and $\gamma$ as the transition width (larger
$\gamma$ means a slower transition between the massive and faint-end
slopes).

We first calculate $\frac{d\log(M_h)}{d\log(M_\ast)}:$

\begin{eqnarray}
\log(M_h) & = & \log(M_1) + \gamma  \log\left[\left(\frac{M_\ast}{M_{\ast,0}}\right)^{\beta/\gamma} + \left(\frac{M_\ast}{M_{\ast,0}}\right)^{\delta/\gamma}\right],\\
\frac{d\log(M_h)}{d\log(M_\ast)} & = & \frac{d\log(M_h)}{dM_\ast} \, \frac{dM_\ast}{d\log(M_\ast)}\\
& = & M_\ast\, \ln(10) \, \frac{d\log(M_h)}{dM_\ast}\\
& = & \frac{\beta \left(\frac{M_\ast}{M_{\ast,0}}\right)^{\beta/\gamma} + \delta \left(\frac{M_\ast}{M_{\ast,0}}\right)^{\delta/\gamma}}{\left(\frac{M_\ast}{M_{\ast,0}}\right)^{\beta/\gamma} + \left(\frac{M_\ast}{M_{\ast,0}}\right)^{\delta/
\gamma}}\\
& = & \beta + (\delta-\beta)\,\left(1+ \left(\frac{M_\ast}{M_{\ast,0}}\right)^\frac{\beta-\delta}{\gamma}\right)^{-1}.
\end{eqnarray}

This justifies the earlier intuition that the functional form for
$M_h(M_\ast)$ transitions between slopes of $\beta$ and $\delta$ with a
width that increases with $\gamma$.  Note that
$\frac{d\log(M_h)}{d\log(M_\ast)}$ is always of order one, as the stellar
mass is always assumed to increase with the halo mass and vice versa
(namely, $\beta>0$ and $\delta>0$).

Next, we approach $\frac{dN}{d\log{M_h}}$.  From analytical results,
we expect a Schechter function for the halo mass function, namely:
\begin{equation}
\frac{dN}{d\log(M_h)} = \phi_0 \, \ln(10) \, \left(\frac{M_h}{M_0}\right)^{1-\alpha}\, \exp\left(-\frac{M_h}{M_0}\right).
\end{equation}
Substituting in the equation for $M_h(M_\ast)$, we have
\begin{eqnarray}
\frac{dN}{d\log(M_h)} & = & \phi_0 \, \ln(10) \, \left[\left(\frac{M_\ast}{M_{\ast,0}}\right)^{\beta/\gamma} + \left(\frac{M_\ast}{M_{\ast,0}}\right)^{\delta/\gamma}\right]^{\gamma(1-\alpha)}\nonumber \\
&&\hspace{-10ex}\times \left(\frac{M_h}{M_0}\right)^{1-\alpha}\, \exp\left(-\frac{M_1}{M_0}\, \left[\left(\frac{M_\ast}{M_{\ast,0}}\right)^{\beta/\gamma} + \left(\frac{M_\ast}{M_{\ast,0}}\right)^{\delta/\gamma}\right]^{\gamma} \right).\hspace{3ex}
\end{eqnarray}

Already evident is the generic result that there will be separate
faint-end and massive-end slopes in the stellar mass function, and
that the falloff is not generically specified by an exponential.  We
may make one simplification in this model---namely, to note that
$M_h(M_{\ast,0})$ corresponds to the halo mass at which the slope of
$M_h(M_\ast)$ begins to transition from $\beta$ to $\delta$.  We
expect this transition to correspond to the transition between
supernova feedback and AGN feedback in semi-analytic models---namely,
for a halo mass which is too large to be affected much by supernova
feedback, but which is yet too small to host a large AGN.  This
implies that $M_h(M_{\ast,0})$ is expected to be around $10^{12}$
$\Msun$ or less, meaning that $M_h/M_0$ is small until stellar
masses well beyond $M_{\ast,0}$, meaning that we may neglect the
faint-end slope of the $M_h(M_\ast)$ relation in the exponential
portion of the stellar mass function:

 \begin{eqnarray}
\frac{dN}{d\log(M_h)} & = & \phi_0 \, \ln(10) \, \left(\frac{M_1}{M_0}\right)^{1-\alpha}\, \left[\left(\frac{M_\ast}{M_{\ast,0}}\right)^{\beta/\gamma} + \left(\frac{M_\ast}{M_{\ast,0}}\right)^{\delta/\gamma}\right]^{\gamma(1-\alpha)}\nonumber \\
&&\times \exp\left(-\frac{M_1}{M_0}\, \left(\frac{M_\ast}{M_{\ast,0}}\right)^{\delta} \right).
\end{eqnarray}

Hence, we may combine these two equations to obtain the expression for
the stellar mass function:
 \begin{eqnarray}
\frac{dN}{d\log(M_\ast)} & = & \phi_0 \, \ln(10) \, \left(\frac{M_1}{M_0}\right)^{1-\alpha}\, \left[\left(\frac{M_\ast}{M_{\ast,0}}\right)^{\beta/\gamma} + \left(\frac{M_\ast}{M_{\ast,0}}\right)^{\delta/\gamma}\right]^{\gamma(1-\alpha)}\nonumber \\
&&\times \exp\left(-\frac{M_1}{M_0}\, \left(\frac{M_\ast}{M_{\ast,0}}\right)^{\delta} \right) \nonumber \\
 && \times \left[ \beta + (\delta-\beta)\,\left(1+ \left(\frac{M_\ast}{M_{\ast,0}}\right)^\frac{\beta-\delta}{\gamma}\right)^{-1}\right].
 \label{smf}
\end{eqnarray}

While this seems complicated, it may be intuitively deconstructed as:
 \begin{eqnarray}
\frac{dN}{d\log(M_\ast)} & = & \left[\mathrm{constant}\right] \, \left[\textrm{double power law}\right] \nonumber \\
&& \times \left[\textrm{exponential dropoff}\right] \, \mathcal{O}(1).
\end{eqnarray}

As mentioned previously, this functional form is equivalent to $\pdirect$.  To convert to the true stellar mass function $\ptrue$ or the observed stellar mass function $\pobs$, it must be convolved with the scatter in stellar mass at fixed halo mass and (for $\pobs$) the scatter in calculated stellar mass at fixed true stellar mass.  As such, it should be clear that---while the final form may be Schechter--\textit{like}---there is certainly much more flexibility in the final shape of the GSMF than a Schechter function alone would allow, as evidenced by the five parameters required to fully specify equation \ref{smf}.

\section{Data Tables}
\label{a_data}

We reproduce here listings of the data points in Figures \ref{smmr_together}, \ref{smmr_together_n}, and \ref{all_cen_together_z4_00} in Tables \ref{t:smf_lowz_syst}, \ref{t:smf_lowz_stat}, and \ref{t:smf_highz}, respectively.  See sections \ref{smmr_results} and \ref{s:discussion_highz} for details on the data points in each table.

\begin{table*}[h]
\caption{Stellar Mass Fractions For $0<z<1$ Including Full Systematics }
\label{t:smf_lowz_syst}
\begin{center}
\begin{tabular}{cccc}
\hline
\hline
& $z = 0.1$ & $z = 0.5$ & $z = 1.0$\\
$\log_{10}(M_h)$ & $\log_{10}(M_\ast/M_h)$& $\log_{10}(M_\ast/M_h)$& $\log_{10}(M_\ast/M_h)$ \\
\hline
11.00 & $-2.30^{+0.26}_{-0.23}$& & \\
11.25 & $-1.96^{+0.25}_{-0.23}$& $-2.11^{+0.22}_{-0.26}$& \\
11.50 & $-1.67^{+0.24}_{-0.24}$& $-1.84^{+0.22}_{-0.26}$& $-2.01^{+0.25}_{-0.24}$\\
11.75 & $-1.53^{+0.23}_{-0.24}$& $-1.70^{+0.24}_{-0.24}$& $-1.85^{+0.26}_{-0.23}$\\
12.00 & $-1.54^{+0.24}_{-0.24}$& $-1.68^{+0.26}_{-0.23}$& $-1.77^{+0.26}_{-0.23}$\\
12.25 & $-1.62^{+0.24}_{-0.24}$& $-1.72^{+0.26}_{-0.23}$& $-1.77^{+0.25}_{-0.23}$\\
12.50 & $-1.74^{+0.24}_{-0.24}$& $-1.81^{+0.25}_{-0.23}$& $-1.81^{+0.24}_{-0.24}$\\
12.75 & $-1.87^{+0.23}_{-0.25}$& $-1.92^{+0.25}_{-0.23}$& $-1.89^{+0.23}_{-0.26}$\\
13.00 & $-2.02^{+0.23}_{-0.25}$& $-2.05^{+0.24}_{-0.24}$& $-1.99^{+0.22}_{-0.27}$\\
13.25 & $-2.18^{+0.22}_{-0.26}$& $-2.19^{+0.24}_{-0.25}$& $-2.11^{+0.21}_{-0.28}$\\
13.50 & $-2.35^{+0.22}_{-0.26}$& $-2.34^{+0.23}_{-0.25}$& $-2.25^{+0.21}_{-0.29}$\\
13.75 & $-2.52^{+0.22}_{-0.27}$& $-2.51^{+0.23}_{-0.26}$& $-2.39^{+0.21}_{-0.30}$\\
14.00 & $-2.70^{+0.21}_{-0.28}$& $-2.68^{+0.23}_{-0.26}$& $-2.55^{+0.22}_{-0.30}$\\
14.25 & $-2.88^{+0.21}_{-0.28}$& $-2.86^{+0.23}_{-0.26}$& \\
14.50 & $-3.07^{+0.20}_{-0.29}$& $-3.04^{+0.23}_{-0.27}$& \\
14.75 & $-3.26^{+0.20}_{-0.30}$& & \\
15.00 & $-3.45^{+0.20}_{-0.30}$& & \\
\hline
\end{tabular}
\\[1ex]
\tablecomments{ Halo masses are in units of $\Msun$.  Constraints are quoted over the mass range probed
by the observed GSMF.}
\end{center}
\end{table*}

\begin{table*}[h]
\caption{Stellar Mass Fractions without Systematic Errors ($\mu = \kappa = 0$)}
\label{t:smf_lowz_stat}
\begin{center}
\begin{tabular}{cccc}
\hline
\hline
& $z = 0.1$ & $z = 0.5$ & $z = 1.0$\\
$\log_{10}(M_h)$ & $\log_{10}(M_\ast/M_h)$& $\log_{10}(M_\ast/M_h)$& $\log_{10}(M_\ast/M_h)$ \\
\hline
11.00 & $-2.30^{+0.03}_{-0.02}$& & \\
11.25 & $-1.96^{+0.04}_{-0.01}$& $-2.10^{+0.04}_{-0.08}$& \\
11.50 & $-1.67^{+0.03}_{-0.01}$& $-1.83^{+0.05}_{-0.06}$& $-2.02^{+0.10}_{-0.06}$\\
11.75 & $-1.53^{+0.01}_{-0.01}$& $-1.71^{+0.06}_{-0.03}$& $-1.85^{+0.10}_{-0.04}$\\
12.00 & $-1.54^{+0.01}_{-0.02}$& $-1.68^{+0.06}_{-0.02}$& $-1.78^{+0.09}_{-0.03}$\\
12.25 & $-1.62^{+0.00}_{-0.02}$& $-1.72^{+0.06}_{-0.01}$& $-1.77^{+0.08}_{-0.03}$\\
12.50 & $-1.74^{+0.01}_{-0.02}$& $-1.81^{+0.05}_{-0.02}$& $-1.81^{+0.06}_{-0.04}$\\
12.75 & $-1.87^{+0.01}_{-0.03}$& $-1.92^{+0.05}_{-0.02}$& $-1.88^{+0.04}_{-0.06}$\\
13.00 & $-2.02^{+0.01}_{-0.03}$& $-2.05^{+0.04}_{-0.03}$& $-1.98^{+0.03}_{-0.08}$\\
13.25 & $-2.18^{+0.02}_{-0.04}$& $-2.19^{+0.04}_{-0.04}$& $-2.10^{+0.04}_{-0.10}$\\
13.50 & $-2.35^{+0.02}_{-0.05}$& $-2.34^{+0.04}_{-0.05}$& $-2.24^{+0.04}_{-0.13}$\\
13.75 & $-2.52^{+0.03}_{-0.06}$& $-2.51^{+0.05}_{-0.06}$& $-2.38^{+0.05}_{-0.15}$\\
14.00 & $-2.70^{+0.03}_{-0.07}$& $-2.68^{+0.05}_{-0.07}$& $-2.54^{+0.06}_{-0.17}$\\
14.25 & $-2.88^{+0.04}_{-0.08}$& $-2.85^{+0.06}_{-0.08}$& \\
14.50 & $-3.07^{+0.04}_{-0.09}$& $-3.04^{+0.06}_{-0.10}$& \\
14.75 & $-3.25^{+0.05}_{-0.10}$& & \\
15.00 & $-3.45^{+0.05}_{-0.11}$& & \\
\hline
\end{tabular}
\\[1ex]
\tablecomments{ Halo masses are in units of $\Msun$.}
\end{center}
\end{table*}

\begin{table*}[h]
\caption{Stellar Mass Fractions For $0.8<z<4$ Including Full Systematics }
\label{t:smf_highz}
\begin{center}
\begin{tabular}{ccccc}
\hline
\hline
& $z = 1.0$ & $z = 2.0$ & $z = 3.0$ & $z = 4.0$\\
$\log_{10}(M_h)$ & $\log_{10}(M_\ast/M_h)$& $\log_{10}(M_\ast/M_h)$& $\log_{10}(M_\ast/M_h)$& $\log_{10}(M_\ast/M_h)$ \\
\hline
11.50 & $-2.01^{+0.25}_{-0.24}$& & & \\
11.75 & $-1.85^{+0.26}_{-0.23}$& & & \\
12.00 & $-1.77^{+0.26}_{-0.23}$& $-1.89^{+0.22}_{-0.27}$& $-1.89^{+0.25}_{-0.27}$& \\
12.25 & $-1.77^{+0.25}_{-0.23}$& $-1.76^{+0.24}_{-0.25}$& $-1.67^{+0.21}_{-0.29}$& $-1.58^{+0.22}_{-0.32}$\\
12.50 & $-1.81^{+0.24}_{-0.24}$& $-1.78^{+0.23}_{-0.26}$& $-1.63^{+0.21}_{-0.30}$& $-1.50^{+0.20}_{-0.35}$\\
12.75 & $-1.89^{+0.23}_{-0.26}$& $-1.87^{+0.20}_{-0.30}$& $-1.71^{+0.19}_{-0.35}$& $-1.56^{+0.19}_{-0.40}$\\
13.00 & $-1.99^{+0.22}_{-0.27}$& $-2.00^{+0.17}_{-0.35}$& $-1.83^{+0.17}_{-0.39}$& $-1.68^{+0.19}_{-0.44}$\\
13.25 & $-2.11^{+0.21}_{-0.28}$& $-2.14^{+0.15}_{-0.39}$& $-1.97^{+0.16}_{-0.44}$& $-1.82^{+0.19}_{-0.49}$\\
13.50 & $-2.25^{+0.21}_{-0.29}$& $-2.30^{+0.15}_{-0.42}$& $-2.13^{+0.17}_{-0.47}$& $-1.98^{+0.19}_{-0.52}$\\
13.75 & $-2.39^{+0.21}_{-0.30}$& $-2.47^{+0.17}_{-0.45}$& $-2.30^{+0.19}_{-0.49}$& \\
14.00 & $-2.55^{+0.22}_{-0.30}$& $-2.64^{+0.20}_{-0.47}$& & \\
\hline
\end{tabular}
\\[1ex]
\tablecomments{ Halo masses are in units of $\Msun$.}
\end{center}
\end{table*}

\end{appendix}

\end{document}